\documentclass[12pt, referee]{aastex}
\usepackage[dvips]{color}

\newcommand{\bk}{\textcolor[rgb]{0,0,0}}

\shorttitle{Cosmic ray ionization rate}
\shortauthors{Neufeld and Wolfire}
\def\gtsima{$\;\buildrel > \over \sim \;$}
\def\simgt{\lower.5ex \hbox{\gtsima}}
\def\ltsima{$\;\buildrel < \over \sim \;$}
\def\simlt{\lower.5ex \hbox{\ltsima}}

\begin{document}

   \title{The cosmic ray ionization rate in the Galactic disk, as determined from observations of molecular ions}
\author{David A. Neufeld\altaffilmark{1} and Mark G. Wolfire\altaffilmark{2}}
\altaffiltext{1}{Department of Physics and Astronomy, Johns Hopkins University,
3400 North Charles Street, Baltimore, MD 21218; neufeld@pha.jhu.edu}
\altaffiltext{2}{Department of Astronomy, University of Maryland, College Park, MD 20742; mwolfire@astro.umd.edu}

  \begin{abstract}
  
  We have obtained estimates for the cosmic-ray ionization rate (CRIR) in the Galactic disk, using a detailed model for the physics and chemistry of diffuse interstellar gas clouds to interpret previously-published measurements of the abundance of four molecular ions: ArH$^+$, OH$^+$, $\rm H_2O^+$ and $\rm H_3^+$.  For diffuse {\it atomic} clouds at Galactocentric distances in the range $R_g \sim 4 - 9$~kpc, observations of ArH$^+$, OH$^+$, and $\rm H_2O^+$ imply a mean primary CRIR of 
  \newline
  $(2.2 \pm 0.3) \exp [(R_0-R_g)/4.7\,\rm{kpc}] \times 10^{-16} \rm \, s^{-1}$ per hydrogen atom, where $R_0=8.5$~kpc.   Within diffuse {\it molecular} clouds observed toward stars in the solar neighborhood, measurements of $\rm H_3^+$ and $\rm H_2$ imply a primary CRIR of $(2.3 \pm 0.6) \times 10^{-16}\,\,\rm s^{-1}$ per H atom, corresponding to a total ionization rate per H$_2$ molecule of $(5.3 \pm 1.1) \times 10^{-16}\,\,\rm s^{-1},$ in good accord with previous estimates.  These estimates are also in good agreement with a rederivation, presented here, of the CRIR implied by recent observations of carbon and hydrogen radio recombination lines along the sight-line to Cas A.  Here, our best-fit estimate for the primary CRIR is $2.9 \times 10^{-16}\,\,\rm s^{-1}$ per H atom.  Our results show marginal evidence that the CRIR in diffuse molecular clouds decreases with cloud extinction, $A_{\rm V}({\rm tot})$, with a best-fit dependence $\propto A_{\rm V}({\rm tot})^{-1}$ for $A_{\rm V}({\rm tot}) \ge 0.5$. 

\end{abstract}
   \keywords{Astrochemistry -- ISM:~molecules -- Submillimeter:~ISM -- Molecular processes -- ISM:~clouds -- cosmic-rays
               }
%

\section{Introduction}

In the century following their discovery by Victor Hess in 1912, cosmic-rays have been recognized as  an important constituent of the Galaxy.   With a total energy density somewhat larger that of starlight (e.g.\ Draine 2011), cosmic-rays are the dominant source of hydrogen 
ionization for the cold neutral medium (CNM)
within the Galactic ISM.  In starless molecular cloud cores, they are also the dominant source of heating.  Thus, cosmic-rays play a central role in astrochemistry by initiating a rich ion-neutral chemistry that operates within the CNM, and the cosmic-ray ionization rate (CRIR) is a key parameter in models of the chemistry of the ISM (Grenier et al.\ 2015, and references therein).  Three related definitions of this parameter are widely used in the literature but must be distinguished.  Here, we adopt the {\it primary} ionization rate per hydrogen atom, $\zeta_p({\rm H})$, as the fundamental parameter of interest, because it is most directly related to the density of cosmic-rays.  The two other quantities of interest are the {\it total} rate of ionization per hydrogen atom,  $\zeta_t({\rm H})$, which includes the secondary ionizations that are caused by the energetic electrons produced by primary ionizations, and the total ionization rate per hydrogen molecule, $\zeta_t({\rm H}_2).$  While the exact ratios of these three quantities depend upon the fractional ionization and molecular fraction (Dalgarno et al.\ 1999), the rough relationship is $\zeta_p({\rm H}) = \zeta_t({\rm H})/1.5 = \zeta_t({\rm H}_2)/2.3 $ under typical conditions within the diffuse neutral ISM (Glassgold \& Langer 1974).

While cosmic-rays of energies above $\sim 1$~GeV can be readily observed from the location of Earth's orbit, cosmic rays of lower energy have their flux modulated by the Sun's magnetic field and the solar wind.  These are precisely the cosmic-rays that dominate the ionization and 
heating of the ISM.  Recent measurements performed by the {\it Voyager I} spacecraft, now located beyond the heliopause, have provided improved estimates of the flux of lower-energy cosmic-ray protons and electrons, down to energies as low as $\sim 3$~MeV (Cummings et al.\ 2016).   Nevertheless, it remains unclear whether the cosmic-ray fluxes reach their unmodulated values even at the current location of {\it Voyager I}.  Moreover, because the ionization cross-sections for H and H$_2$ peak at an energy $\sim 0.01~$MeV, an extrapolation to unobserved energies is still required to determine the implied CRIR, which remains quite uncertain even in the solar neighborhood.

Estimates of the CRIR in interstellar gas clouds may be obtained through a careful astrochemical analysis of the observed abundances of specific molecules whose production is driven by cosmic-rays.
In dense molecular clouds that are shielded from the interstellar UV radiation field, the abundances of H$^{13}$CO$^+$ and H$_3^+$ have been used to derive estimates of $\zeta_t({\rm H}_2)$
in the range $\sim 0.6$ to $6 \times 10^{-17}\, \rm s^{-1}$ (van der Tak \& van Dishoeck 2000; Kulesa 2002).   For the cold diffuse interstellar medium, where the UV radiation field is less strongly attenuated, it is convenient to use the nomenclature adopted by Snow \& McCall (2006), who distinguished between diffuse {\it atomic} material, in which the molecular fraction $f_{\rm H2} = 2\,n({\rm H}_2)/[2\,n({\rm H}_2)+n({\rm H})]$ is smaller than 0.1, and diffuse {\it molecular} material, in which $f_{\rm H2}$ is larger than 0.1 but the UV radiation field is still sufficient to maintain C$^+$ as the dominant reservoir of gas-phase carbon nuclei.  Diffuse {\it atomic} gas is found in clouds of typical visual extinction $A_{\rm V} \le 0.2$~mag and H nucleus density $n_{\rm H} = 10 - 100\, \rm cm^{-3},$ while diffuse {\it molecular} gas is found in clouds with $A_{\rm V} =  0.2 - 1$~mag and $n_{\rm H} = 100 - 500\, \rm cm^{-3}$ (Snow \& McCall 2006). Clearly, the distinction here -- although useful -- is somewhat arbitrary, and we certainly expect a continuous distribution of $f_{\rm H2}$, $A_{\rm V}$ and $n_{\rm H}$.

The CRIR within diffuse {\it molecular} gas can be inferred from measurements of H$_3^+$ (e.g.\ Indriolo et al.\ 2007) and HD (e.g.\ Liszt 2015).   
Such measurements have revealed that the CRIR within the diffuse molecular clouds is typically an order-of-magnitude larger than those inferred for dense molecular clouds, suggesting that the cosmic-ray fluxes are significantly attenuated within dense molecular clouds.  These measurements generalize the surprising result obtained in the pioneering study of McCall et al.\ (2003), which combined astronomical observations with a new laboratory measurements of the dissociative recombination rate for H$_3^+$ and derived a CRIR along the sight-line to $\zeta$~Per that was a factor $\sim 40$ larger than those typically inferred for dense molecular clouds.
Similarly-enhanced CRIRs were subsequently inferred from measurements of the OH$^+$ and H$_2$O$^+$ abundances in the diffuse ISM (e.g.\ Gerin et al.\ 2010; Neufeld et al.\ 2010); in this case, the molecular fraction indicated by the OH$^+$/H$_2$O$^+$ column density ratio is $\sim 2 - 10\,\%$ (Indriolo et al.\ 2015, hereafter I15), implying that the CRIR estimates obtained from measurements of OH$^+$ and H$_2$O$^+$ apply to diffuse {\it atomic} material.

The past five years have seen the publication of two large surveys of relevance to the CRIR in the diffuse ISM: a near-infrared survey of H$_3^+$ in diffuse molecular clouds, obtained with ground-based telescopes (Indriolo \& McCall 2012; hereafter IM12); and a submillimeter survey of   
OH$^+$ and H$_2$O$^+$ in diffuse atomic clouds (I15), obtained using the {\it Herschel Space Observatory}.
As will be discussed in Sections 3 and 4 below, these studies used simple analytic expressions -- based upon an approximate treatment of the chemistry -- to estimate CRIRs from the observed abundance of H$_3^+$ or the observed abundances of OH$^+$ and H$_2$O$^+$.   

In this paper, we will present the results of detailed physical and chemical models for diffuse interstellar gas clouds, examine critically the approximations used by IM12 and I15, and present refined estimates for the CRIR in the diffuse ISM.  The diffuse cloud model used in this study is described in \S 2, along with the  $\rm H_3^+$ 
abundance predictions obtained from the model.  In \S 3, we present estimates of the CRIR 
within diffuse molecular and diffuse atomic clouds in the Galactic disk.  
In \S 4 we discuss the comparison with previous estimates reported in the literature and present recommended values for the mean CRIR in the Galactic disk.

\section{Diffuse cloud model, and predictions for $\rm H_3^+$ abundances}

\subsection{Diffuse cloud model}

Our thermal and chemical model for diffuse molecular clouds is based on that described by Hollenbach et al. (2012; hereafter H12), with the modifications
discussed by Neufeld \& Wolfire (2016; hereafter Paper I).  In this model,
we treat an interstellar gas cloud as a two-sided slab that is illuminated isotropically by an ultraviolet radiation field with the spectrum given by Draine (1978).  The strength of the radiation field is characterized by the quantity, $\chi_{\rm UV}$, which is the ratio of the specific intensity to the mean interstellar value recommended by Draine (1978).  The attenuation of the isotropic field was
calculated as described in Wolfire et al.\ (2010),
and the equilibrium gas temperature and steady-state chemical abundances 
were calculated as a function of depth into the cloud.  As in Paper I, we included a network of chemical reactions for argon-containing species identical to what we presented in Schilke et al.\ (2014; hereafter S14).

\subsection{Standard model grid}

\begin{deluxetable}{lll}
\tablewidth{0pt}
\tablecaption{Grid of model parameters}
\tablehead{Parameter & Number of values & Values}
\startdata
$\chi_{\rm UV}$ & 10 & 0.05, 0.1, 0.2, 0.3, 0.5, 1.0, 2.0, 3.0, 5.0, 10.0 \\
$\zeta_p({\rm H})/10^{-16}\,\rm s^{-1}$ & 9 & 0.006, 0.02, 0.06, 0.2, 0.6, 2.0, 6.0, 20, 60 \\ 
$A_{\rm V}({\rm tot})$/mag & 16 & 0.0003, 0.001, 0.003, 0.01, 0.03, 0.1, 0.2, \\
& & 0.3, 0.5, 0.8, 1.0, 1.5, 2.0, 3.0, 5.0, 8.0 \\
$Z/Z_{\rm std}$ & 2 & 1.0, 2.0 \\
$n_{\rm H}$ & 1 & 50$\,\rm cm^{-3}$ \\
\enddata
\end{deluxetable}

We have computed a grid of models for diffuse atomic clouds, and for diffuse and translucent molecular clouds, for all combinations listed in Table 1 of four key parameters: the normalized UV radiation field, $\chi_{\rm UV}$, the primary CRIR per H atom, $\zeta_p({\rm H})$, the total visual extinction across the slab, $A_{\rm V}({\rm tot})$, and the metallicity, $Z$.  In our standard metallicity model, $Z=Z_{\rm std}$, the adopted elemental abundances were those most appropriate to the Galactic ISM at the solar circle, 
for which we assumed gas-phase carbon, oxygen and argon abundances of  $1.6 \times 10^{-4}$ (Sofia et al. 2004; Gerin et al.\ 2015), $3.9 \times 10^{-4}$ (Cartledge et al. 2004), and $3.2 \times 10^{-6}$ (Asplund et al.\ 2009) respectively relative to H nuclei.
Given a primary CRIR per H atom, $\zeta_p({\rm H})$, we used the expressions given by Dalgarno et al.\ (1999) to determine the total CRIR per H atom (including the effects of secondary ionizations), $\zeta_t({\rm H})$, and the total ionization rate per H$_2$ molecule, $\zeta_t({\rm H}_2)$.  The typical conversion factors are in good agreement with those given 
by Glassgold \& Langer (1974): $\zeta_t({\rm H})= 1.5 \zeta_p({\rm H})$ and $\zeta_t({\rm H}_2)= 2.3 \zeta_p({\rm H})$

All the models were computed for a single H nucleus density, $n_{\rm H} = 50\,\rm cm^{-3}$; however, as  explained in Paper~I, the cloud properties can be predicted for other values of $n_{\rm H}$ by means of a simple scaling, because the cloud properties are completely determined by $\chi_{\rm UV}/n_{\rm H}$, $\zeta_p({\rm H})/n_{\rm H}$, $A_{\rm V}({\rm tot})$, and $Z$.  The selection of parameters listed in Table 1 extends the range of those considered in Paper I to smaller $\chi_{\rm UV}$, to smaller $\zeta_p({\rm H})$, and to larger $A_{\rm V}({\rm tot})$, resulting in a grid consisting of 2880 diffuse cloud models.

\subsection{Predictions for the $\rm H_3^+$ abundance}

Our treatment of the chemistry of $\rm OH^+$, $\rm H_2O^+$ and $\rm ArH^+$ has been presented in previous papers (H12, S14, Paper I) and will not be discussed further here.  In this section, we confine our attention to the H$_3^+$ molecular ion.  
As has been described in many previous studies (e.g. IM12 and references therein), the formation of H$_3^+$ is initiated by the cosmic ray ionization of H$_2$ to form H$_2^+$, followed by proton transfer from to H$_2$:
$$\rm H_2^+ + H_2 \rightarrow H_3^+ + H. \eqno(R1)$$
If the molecular fraction is small, charge transfer with H is a significant competing channel that limits the H$_3^+$ abundance:
$$\rm H_2^+ + H \rightarrow H^+ + H_2. \eqno(R2)$$
Dissociative recombination and photoionization are other loss processes for $\rm H_2^+$, but they are generally unimportant, so that the fraction of ${\rm H}_2$ ionizations that are followed by H$_3^+$ production, $\epsilon({\rm H}_3^+)$, is well-approximated by 
$$\epsilon({\rm H}_3^+)= {1 \over 1 + k_2 n({\rm H})/k_1 n({\rm H_2})}=
{1 \over 1 + 0.3\, n({\rm H})/n({\rm H_2})},\eqno(1)$$ 
where $k_1$ and $k_2$ are the rate coefficients for reactions (R1) and (R2) respectively, for which we adopt values of $2.1 \times 10^{-9}$ (Theard \& Huntress 1974) and $6.4 \times 10^{-10}\, \rm cm^3\,s^{-1}$ (Karpas \& Huntress 1979).  The H$_3^+$ production efficiency, $\epsilon({\rm H}_3^+)$, exceeds 50$\%$ whenever $n({\rm H})/n({\rm H_2})$ is smaller than $\sim 3$, or equivalently whenever the molecular fraction,  $f({\rm H}_2) = 2n({\rm H}_2)/[n({\rm H})+2n({\rm H}_2)]$, exceeds $\sim 0.4$.

Except in dense clouds, where the electron fraction $x_{\rm e}$ is very small, 
the destruction of H$_3^+$ is dominated by dissociative recombination:
$$\rm H_3^+ + e \rightarrow H_2 + H\,\,\,\,or\,\,\,\,3H. \eqno(R3)$$  At small visual 
extinctions, carbon is fully ionized; 
$x_{\rm e}$ is at therefore least\footnote{For sufficiently large CRIRs, as discussed further below, the ionization of hydrogen can be a significant source (or even the dominant source) of electrons.} as large as the gas-phase abundance of carbon nuclei, $x_{\rm C}$, for which we adopt a value of $1.6 \times 10^{-4}$ (Sofia et al.\ 2004) in our standard metallicity models.    As the  extinction increases, the ionized fraction for carbon begins to drop, and the destruction rate decreases accordingly.  Eventually, for sufficiently small electron fractions, proton transfer to neutral species (such as O or CO) becomesß the dominant loss process and sets a floor on the destruction rate for H$_3^+$.

In diffuse and translucent clouds, where dissociative recombination dominates the destruction of H$_3^+$, the equilibrium $n({\rm H_3^+})/n({\rm H_2})$ density ratio is therefore
$${n({\rm H_3^+}) \over n({\rm H_2})} = {\epsilon({\rm H}_3^+) \, \zeta_t({\rm H}_2) \over k_3 x_e n_{\rm H}},\eqno(2)$$
where $k_3$ is the rate coefficient for reaction (R3) (including both reaction channels).
Given the value of $k_3$ measured by McCall et al.\ (2004), which may be approximated by 
$1.2 \times 10^{-7} (T/100\,{\rm K})^{-0.5},$ we obtain
$${n({\rm H_3^+})\over n({\rm H_2})} = 2.1 \times 10^{-7}\, {\epsilon({\rm H}_3^+)\, \zeta_t({\rm H}_2)_{-15}\,T_2^{0.5} \over (x_{\rm e}/x_{\rm C})\, (Z/Z_\odot)\,
 n_{250}}, \eqno(3)$$
where $\zeta_t({\rm H}_2)_{-15}=\zeta_t({\rm H}_2)/[10^{-15}\,\rm s^{-1}]$, $T_2=T/
[\rm 100\,K]$,
$n_{250}=n_{\rm H}/[250\,\rm cm^{-3}],$ and $Z$ is the metallicity (with $Z/Z_\odot$ = 1 in the standard metallicity case and 2 in the enhanced metallicity case.)  

\begin{figure}
\includegraphics[width=13 cm]{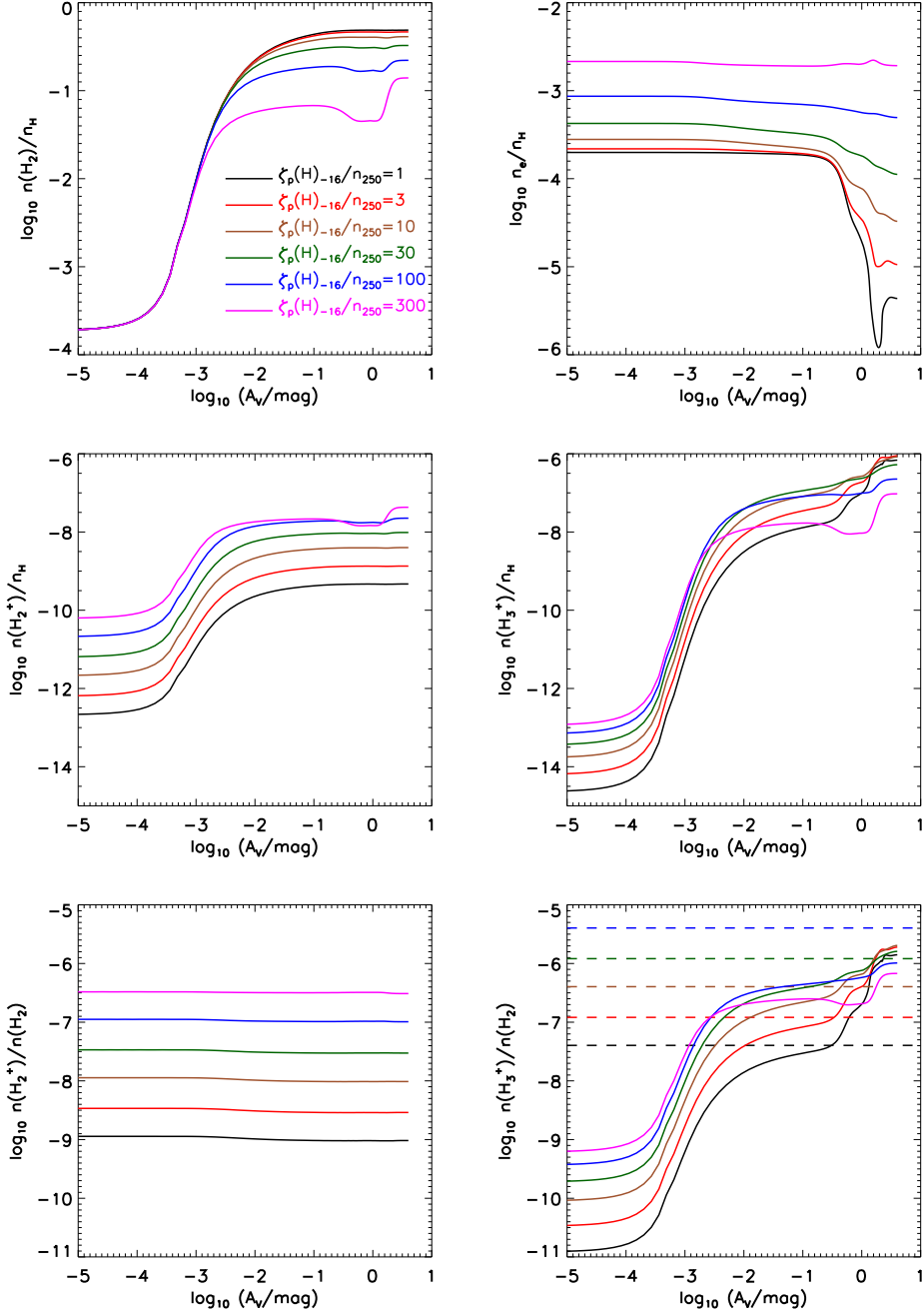}
\caption{Local abundances of H$_2$, electrons, H$_2^+$ and 
H$_3^+$ relative to H nuclei, and of
H$_2^+$ and H$_3^+$ relative to H$_2$, as a function of depth into a cloud of total visual extinction $A_{\rm V}({\rm tot}) = 8$ (i.e.\ $A_{\rm V}({\rm tot}) = 4$ to the midplane)
exposed to UV radiation
with $\chi_{\rm UV}/n_{250} = 1.$ Results are shown for several values of the CRIR
(see legend in top left panel). In the lower right panel, which shows 
$N({\rm H}_3^+)/N({\rm H}_2)$,
the dashed lines indicate predictions of the analytic treatment adopted by IM12
(see our eqn.\ 4), for an assumed gas temperature of 70 K.ß}
\end{figure}

In Figure 1, we have plotted several profiles showing the dependence predicted by our model for
various abundances, 
as a function of depth into a cloud of total visual extinction
$A_{\rm V}({\rm tot}) = 8$ exposed to UV radiation
with $\chi_{\rm UV}/n_{250} = 1.$  Results are shown for six different CRIRs: $\zeta_p({\rm H})_{-16}/n_{250} = 1$ (black), 3 (red), 10 (brown), 30 (green), 100 (blue), and 300 (magenta), where $\zeta_p({\rm H})_{-16} = \zeta_p({\rm H})/[10^{-16}\,\rm s^{-1}] \sim 4.3 \,\zeta_t({\rm H}_2)_{-15}$.  The top left panel, which shows the abundance of H$_2$ relative to H nuclei, reveals a strong gradient resulting from the effect of self-shielding on the H$_2$ photodissociation rate .  In the cloud interior, destruction of H$_2$ by cosmic-rays reduces the H$_2$ abundance by a factor greater than 2 if $\zeta_p({\rm H})_{-16}/n_{250}$ exceeds $\sim 50$ (blue and magenta curves.)  The top right panel shows the electron abundance, $x_{\rm e} = 
n_{\rm e}/n_{\rm H}.$  As discussed previously, significant (i.e.\ greater than a factor 2) departures from $x_{\rm e} = x_{\rm C}$ are predicted either (1) if the CRIR $\zeta_p({\rm H})_{-16}/n_{250}$ exceeds $\sim 50$, in which case the ionization of H by CR can significantly enhance the electron abundance {\it above} $x_{\rm C}$; or (2) once $A_{\rm V}$ exceeds $\sim 0.3\,\rm mag$, at which point the C$^+$ abundance starts to fall (and thus the electron abundance drops {\it below} $x_{\rm C}$ unless condition (1) also applies.) 

The middle panels show the H$_2^+$ and H$_3^+$ abundances relative to H nuclei, while the bottom panels show the H$_2^+$ and H$_3^+$ abundances relative to H$_2$ molecules.  The $n({\rm H}_2^+)/n({\rm H}_2)$ ratio (bottom left panel) is exactly proportional to the CRIR  and shows only a weak dependence on $A_{\rm V}$: a small decrease in $n({\rm H}_2^+)/n({\rm H}_2)$ occurs as the gas becomes molecular, at $A_{\rm V} \sim 10^{-2}$, because the destruction rate in fully-molecular gas, $k_1 n({\rm H}_2) =  k_1 n_H /2$, is somewhat larger than that in fully-atomic gas, $k_2 n({\rm H}) = k_2 n_H$.  The $n({\rm H}_3^+)/n({\rm H}_2)$ ratio (bottom right panel) shows a more complicated behavior.   In their derivation of CRIRs from the observed column densities of $\rm H_3^+$ and $\rm H_2$, IM12 made the simplifying assumptions $x_{\rm e} = x_{\it C}$ and $\epsilon({\rm H}_3^+) = 1$. For a temperature of 70~K, the default value assumed by IM12 unless an alternative estimate was available, and with the approximation that $\zeta_p({\rm H})_{-16} = 4.3 \,\zeta_t({\rm H}_2)_{-15}$, these assumptions then imply 
$${n({\rm H_3^+}) \over n({\rm H_2})} = 4.1 \times 10^{-8}\, {\zeta_p({\rm H})_{-16} \over n_{250}}. \eqno(4)$$   That value is shown by the horizontal dashed lines in the bottom right panel of Figure 1 (with the same color-coding as the solid curves).  Clearly, while  equation (4) provides an adequate description over part of the relevant parameter space, significant deviations do result from shortcomings in the assumption that 
$x_{\rm C} = x_{\rm e}$ (described above), and from departures from $\epsilon({\rm H}_3^+) = 1$ that are important when the molecular fraction is small (see eqn.\ 1 above).

\begin{figure}
\includegraphics[width=15 cm]{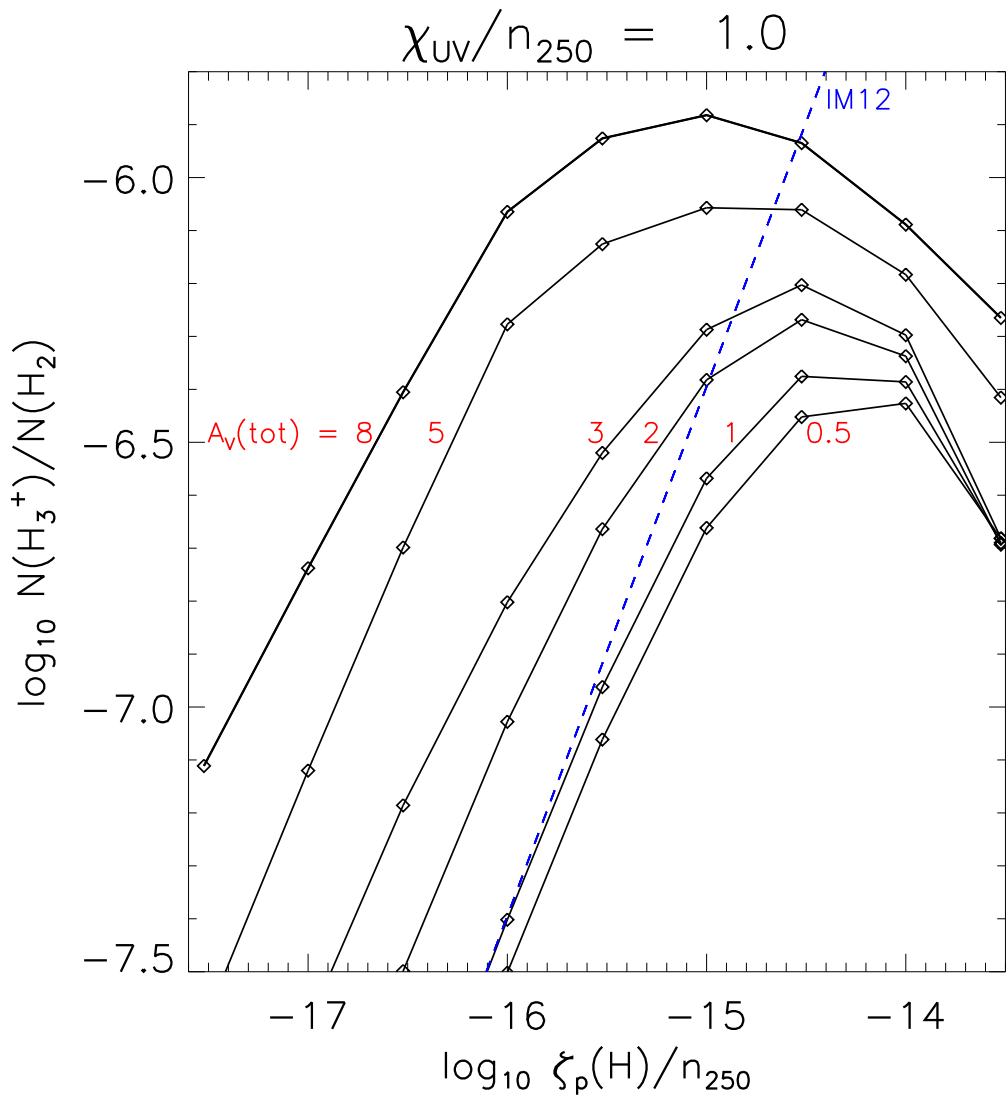}
\caption{$N({\rm H}_3^+)/N({\rm H}_2)$
column density ratios predicted for diffuse and translucent
molecular clouds exposed to UV radiation
with $\chi_{\rm UV}/n_{250} = 1.$  Results are shown for several values of the total visual extinction through the cloud.
The blue dashed line indicates results obtained using 
the analytic treatment adopted by IM12
(see our eqn.\ 4), for an assumed gas temperature of 70 K.}
\end{figure}

Molecular ``abundances" determined from astronomical observations are typically the ratios of {\it column densities}, not number densities.  Accordingly, it is most useful to provide predictions for  ${N({\rm H_3^+}) / N({\rm H_2})}$ along a sightline passing through an interstellar gas cloud.  These are shown in Figure 2, where we have plotted ${N({\rm H_3^+}) / N({\rm H_2})}$ as a function of ${\zeta_p({\rm H})_{-16}/ n_{250}}$.  Results are shown for several values of the total extinction, $A_{\rm V}({\rm tot})$, but they all apply to $\chi_{\rm UV}/n_{250} = 1.$   Here, we also show the approximate results obtained using 
equation (4) (blue dotted line), upon which the CRIR-determinations of IM12 were based. 
As expected from equation (4), ${N({\rm H_3^+}) / N({\rm H_2})}$ initially shows a linear dependence upon the CRIR.  However, once the electron abundance starts to rise above the gas-phase elemental abundance of carbon, the ${N({\rm H_3^+}) / N({\rm H_2})}$ ratio then flattens out.  For the highest CRIRs that we considered, the ${N({\rm H_3^+}) / N({\rm H_2})}$ eventually becomes a decreasing function of the CRIR, because the atomic hydrogen abundance increases sufficiently to compete with H$_2$ for H$_2^+$, reducing $\epsilon({\rm H}_3^+)$ even at the cloud center.  As a result, the ${N({\rm H_3^+}) / N({\rm H_2})}$ ratio is a non-monotonic function of the CRIR.  Figure 2 also shows that the ${N({\rm H_3^+}) / N({\rm H_2})}$ ratio is an increasing function of $A_{\rm V}({\rm tot})$ (even though expression (4) predicts no such dependence).  This behavior occurs because the C$^+$ abundance in the cloud interior is smaller in clouds of larger $A_{\rm V}({\rm tot})$.

\section{Estimates of the CRIR}

\subsection{Diffuse molecular clouds}

\begin{figure}
\includegraphics[width=14 cm]{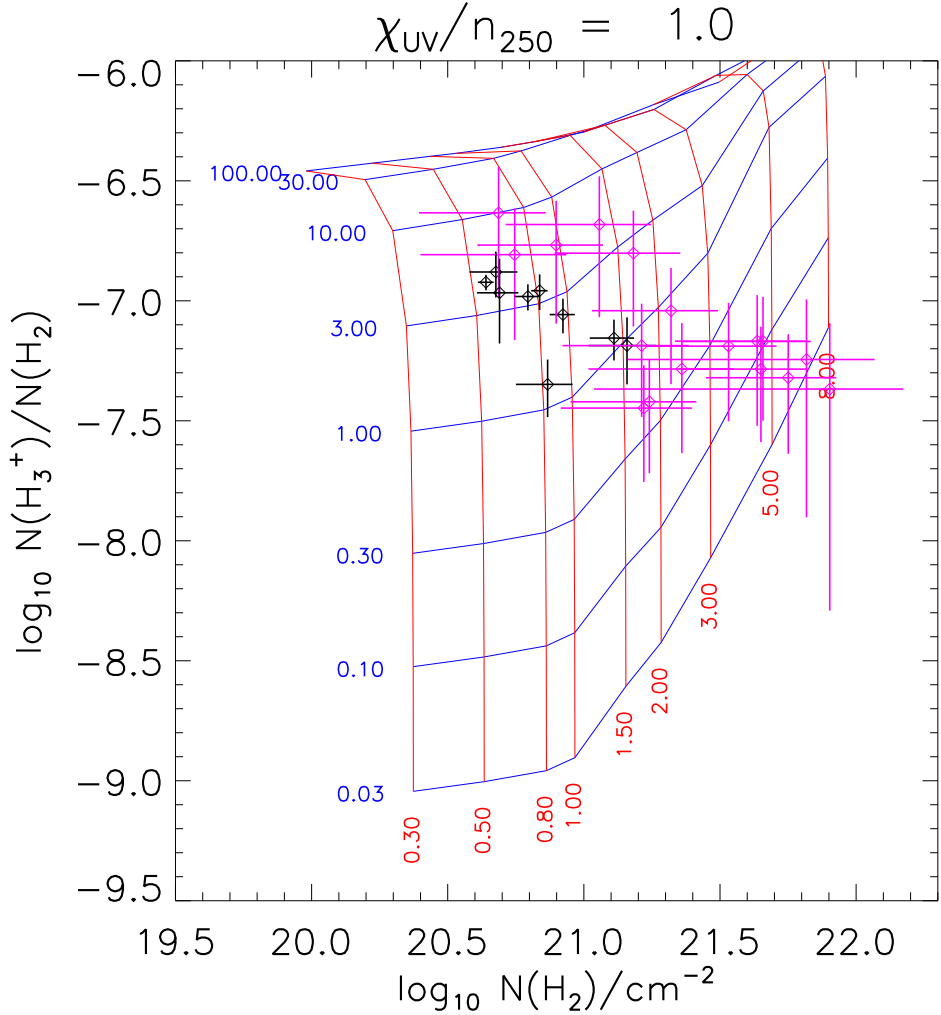}
\caption{H$_2$ column densities and ${N({\rm H_3^+}) / N({\rm H_2})}$ column density 
ratios predicted for
diffuse and translucent molecular clouds with $\chi_{\rm UV}/n_{250} = 1$, 
where $\chi_{\rm UV}$ is the incident
radiation field in Draine (1978) units and $n_{\rm H} = 250\,n_{250}\,\rm cm^{-3}$
is the density of H
nuclei. Results are shown in the plane of $N({\rm H_2})$ and 
${N({\rm H_3^+}) / N({\rm H_2})}$, 
with contours of
visual extinction, $A_{\rm V}({\rm tot})$, 
shown in red and contours of $\zeta_p({\rm H})/n_{250}$ shown in blue
(where $\zeta_p({\rm H}) \sim \zeta_t({\rm H}_2)/2.3$ is the primary cosmic-ray ionization rate per H nucleus
and $\zeta_t({\rm H}_2)$ is the total cosmic-ray ionization rate per H$_2$ molecule.)
Blue contours are labeled with $\zeta_p({\rm H})/n_{250}$, in units of $10^{-16}\,\rm s^{-1}$, and red contours with $A_{\rm V}({\rm tot})$ in mag.  
Diamonds, with 1$\sigma$ error bars, indicate measurements reported by
IM12 or Albertsson et al.\ (2014). Here, black diamonds denote measurements 
obtained from direct observations of
H$_2$, with magenta diamonds showing cases in which the H$_2$ column densities have
been inferred indirectly from observations of CH or $E(B-V)$.}
\end{figure}

For diffuse molecular clouds, IM12 have presented an extensive compilation of 
H$_3^+$ column densities derived from near-IR spectroscopy of stars.  This compilation, presented in their Table 4, includes 21 sight-lines with H$_3^+$ detections, 10 of which had been reported previously, and 30 sight-lines with upper limits.  Two of the sight-lines with H$_3^+$ detections exhibit two (velocity-resolved) absorption components, leading to a total of 23 clouds in which $N({\rm H}_3^+)$ has been measured.  For six of these diffuse clouds, $N({\rm H}_2)$ has been measured directly by means of ultraviolet absorption line spectroscopy.  For the remaining 17 clouds with ${\rm H}_3^+$ detections, 
direct measurements of H$_2$ were unavailable, and IM12 inferred $N({\rm H}_2)$ indirectly from measurements of the selective extinction, $E(B-V)$, or the CH column density.  For these clouds without direct measurements of H$_2$, the inferred H$_2$ column densities were relatively inaccurate, with estimated uncertainties of a factor 2 (i.e.\ 0.30 dex, for those derived from $E(B-V)$) and 1.6 (i.e.\ 0.21 dex, for those derived from $N({\rm CH})).$  In the analysis presented below, we will focus primarily on ``gold-standard" determinations in which $N({\rm H}_3^+)$ and $N({\rm H}_2)$ have been measured directly.  Three more such determinations, reported by Albertsson et al.\ (2014, herefater A14), may be added to the six cases presented by IM12, for a total of 9 clouds with direct measurements of $N({\rm H}_3^+)$ and $N({\rm H}_2).$ 

\begin{figure}
\includegraphics[width=14 cm]{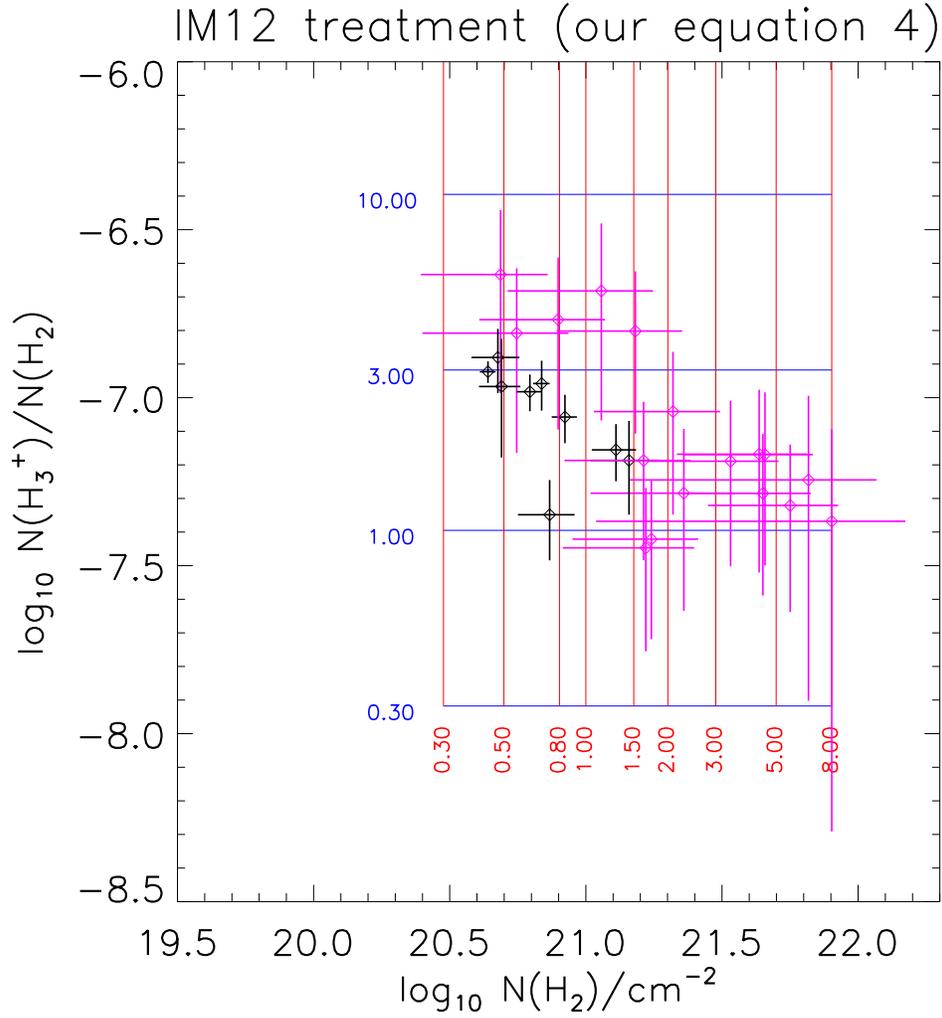}
\caption{Same as Figure 3, but with the column densities
computed using the simple analytic approximations adopted by IM12. Here, we
adopted a gas temperature of 70 K and assumed a molecular fraction of 1.0}
\end{figure}

In Figure 3, we show the data presented by IM12 and A14 in the plane of observables, with
the horizontal axis showing the H$_2$ column density and the vertical axis the column density ratio, $N({\rm H}_3^+)/N({\rm H}_2).$  Here, black diamonds with 1 $\sigma$ error bars refer to clouds with direct measurements of both $N({\rm H}_3^+)$ and $N({\rm H}_2),$ while magenta diamonds refer to clouds for which the H$_2$ column density was estimated from $N({\rm CH})$ or $E(B-V)$.  Overplotted contours show the predictions of our diffuse cloud model, with contours of visual extinction shown in blue, and contours of CRIR shown in red.  Blue contours are labeled with $\zeta_p({\rm H})/n_{250}$ in units of $10^{-16}\,\rm s^{-1}$, and red contours with $A_{\rm V}({\rm tot})$ in mag.  All the model predictions apply to a UV radiation field with $\chi_{UV}/n_{250} = 1.$   The considerations discussed in Section 2.3 are illustrated clearly when this figure is compared with Figure 4, in which identical data are shown with model predictions from the simple analytic treatment of IM12.  In Figure 4, the blue contours of constant CRIR are horizontal and evenly spaced, because the predicted $N({\rm H}_3^+)/N({\rm H}_2)$ ratio is simply proportional to $\zeta_p({\rm H})/n_{250}$.  In Figure 3, by contrast, the blue contours curve upwards at large $N({\rm H}_2)$ because the abundance of electrons -- which destroy H$_3^+$ -- is smaller at larger visual extinctions where carbon is no longer fully ionized.  For larger CRIRs, the spacing between the blue contours diminishes -- and the contours may even cross -- because cosmic-ray ionization of H enhances the electron abundance.  Moreover, the red contours curve to the left near the top of Figure 3, owing to the destruction of H$_2$ by cosmic-rays.

\begin{figure}
\includegraphics[width=14 cm]{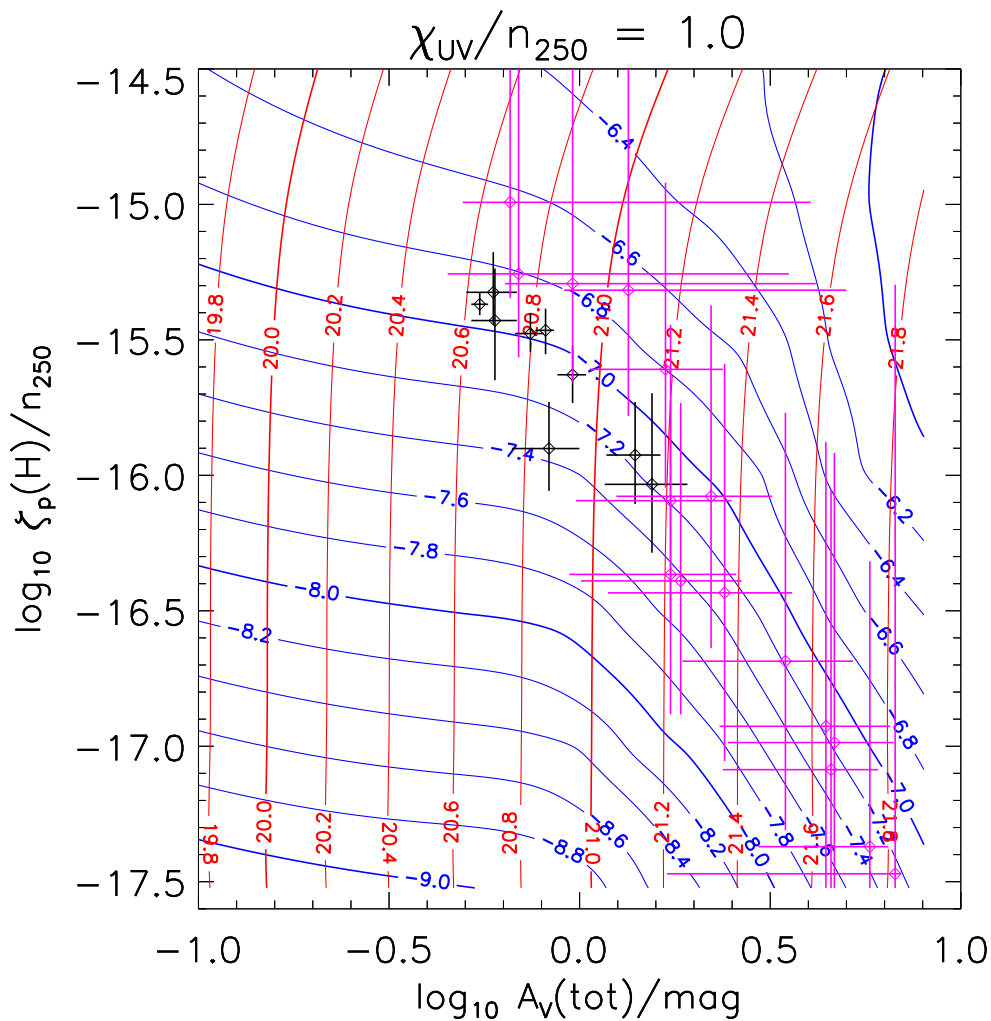}
\caption{
Same as Figure 3, but now with contours of the 
observed quantities $N({\rm H}_2)$ and
$N({\rm H}_3^+)/N({\rm H}_2)$ in the plane of the model parameters 
$A_{\rm V}({\rm tot})$ and $\zeta_p({\rm H})/n_{250}$.
Blue contours are labeled with ${\rm log}_{10}[N({\rm H}_3^+)/N({\rm H}_2)]$, and red contours are labeled with ${\rm log}_{10}[N({\rm H}_2)/{\rm cm}^{-2}]$.
}
\end{figure}

\begin{figure}
\includegraphics[width=14 cm]{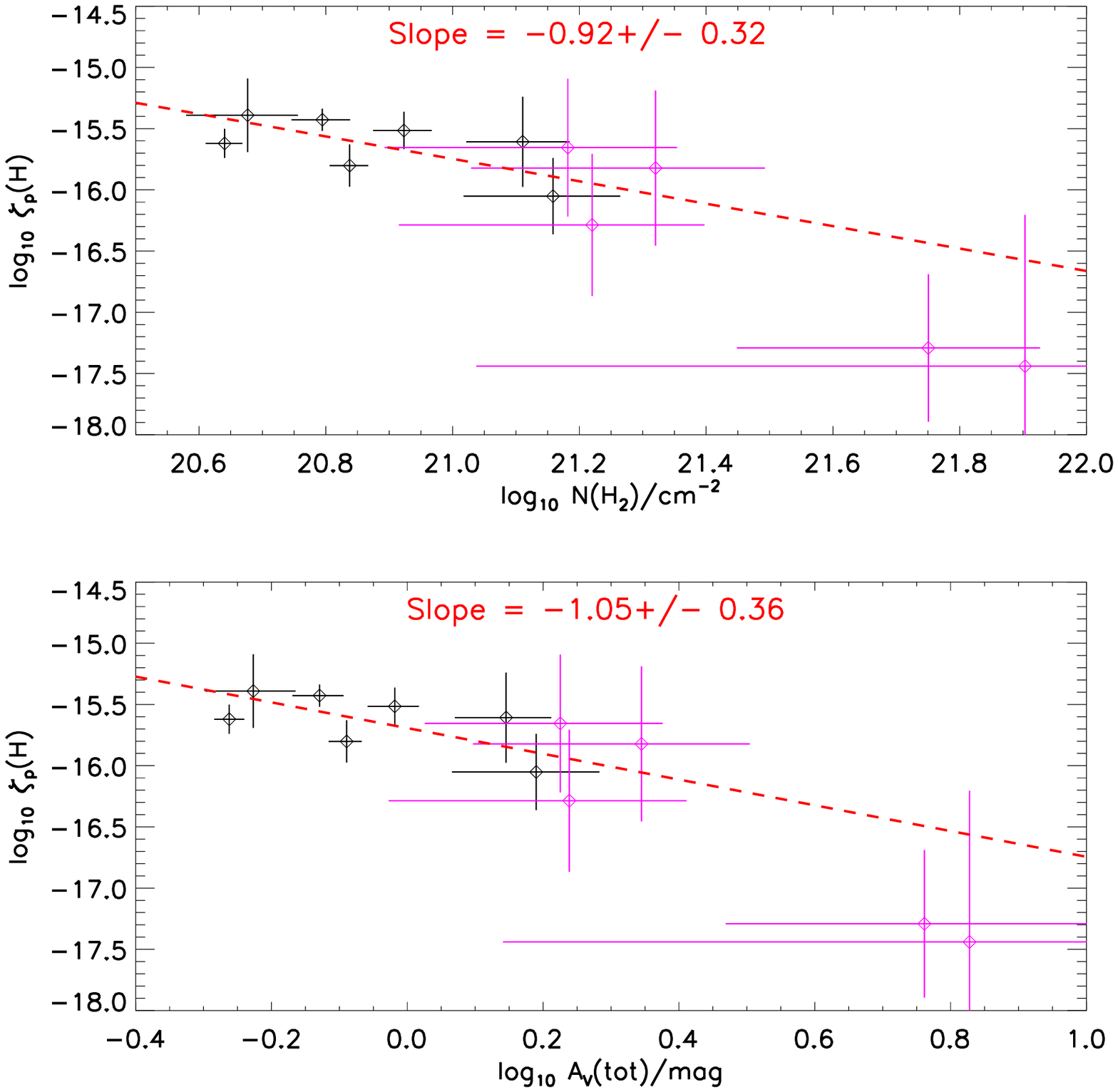}
\caption{
Estimates of $\zeta_p({\rm H})$, derived from measurements of the $\rm H_2$ and $\rm H_3^+$ column densities, as as a function of the measured $N({\rm H}_2)$ (upper panel) and of the derived $A_{\rm V}({\rm tot})$ (lower panel).  Black diamonds: measurements obtained from direct observations of
H$_2$.  Magenta diamonds: measurements in which the H$_2$ column densities have
been inferred indirectly from observations of CH or $E(B-V)$.
Dashed red lines: best fits to all the data.
}
\end{figure}

In Figure 5, we have transformed the coordinate system adopted in Figures 3 and 4, plotting {\it contours of observable quantities in the plane of model parameters}.  Here, the horizontal axis shows the visual extinction, $A_{\rm V}({\rm tot})$, and the vertical axis shows $\zeta_p({\rm H})/n_{250}$.  Blue contours show the logarithm of the $N({\rm H}_3^+)/N({\rm H}_2)$ ratio, and red contours show the logarithm of $N({\rm H}_2)$ in cm$^{-2}$.  The diamonds now represent the best-fit model parameters for each cloud, and the error bars represent 68$\%$ confidence limits.  While the data plotted here reveal a clear tendency for $\zeta_p({\rm H})/n_{250}$ to decrease with $A_{\rm V}({\rm tot})$, it is not clear from Figure 5 whether this tendency occurs because CRIR is a decreasing function of $A_{\rm V}({\rm tot})$ or because the density increases with $A_{\rm V}({\rm tot})$ (or both).  Certainly, there is an expectation that the typical gas density will increase with $A_{\rm V}({\rm tot})$ once self-gravity becomes important.  

For seven of the nine clouds with direct measurements of both $N({\rm H}_3^+)$ and $N({\rm H}_2),$ and for five additional clouds with indirect determinations of $N({\rm H}_2),$
gas density estimates, $n_{\rm H}$, are also available (Sonnentrucker et al.\ 2007).  These density estimates were inferred from a fit to the relative level populations
of rotational states of the C$_2$ molecule, which had been obtained from absorption-line observations at visible wavelengths.   For these clouds, we have multiplied the $\zeta_p({\rm H})/n_{250}$ estimates derived from the $\rm H_2$ and $\rm H_3^+$ column densities by the gas density estimates presented by Sonnentrucker et al.\ 2007, thereby obtaining estimates of the CRIR.  The results are shown in Figure 6, as a function of the measured $N({\rm H}_2)$ (upper panel) and of the derived $A_{\rm V}({\rm tot})$ (lower panel).  All results were obtained for a UV radiation field $\chi_{UV}/n_{250} = 1.$   As in Figures 4 and 5, black diamonds with 1$\sigma$ error bars refer to clouds with direct measurements of both $N({\rm H}_3^+)$ and $N({\rm H}_2),$ while magenta diamonds refer to clouds for which the H$_2$ column density was estimated from $N({\rm CH})$ or $E(B-V)$.
Dashed red lines in Figure 6 represent the best linear fits to the dependence of ${\rm log}_{10}[\zeta_p({\rm H})]$ on the measured ${\rm log}_{10}[N({\rm H}_2)]$ and on the derived ${\rm log}_{10}[A_{\rm V}({\rm tot})]$.  The best-fit slopes are $\sim -1$, but the differences from zero are only of marginal significance.  

\subsection{Diffuse atomic clouds}

As discussed in H12, S14, and Paper I, the CRIR in diffuse {\it atomic} clouds may be probed using observations of $\rm OH^+$, $\rm H_2O^+$ and $\rm ArH^+$.  Model predictions for $\rm ArH^+$ were presented previously in Paper I (their Figure 3) , and those for $\rm OH^+$ and $\rm H_2O^+$ in H12 (their Figures 14 and 15).  In the present study, our results for $\rm OH^+$ and $\rm H_2O^+$ reflect several changes to the chemistry described in Paper I, and have been computed on a finer grid that those presented in H12.  Accordingly, we have shown the results of the current model in Figures 7 -- 9, in a manner analogous to that adopted for Figures 3 -- 5.

In Figure 7, as in H12 and Figure 3 above, we show model predictions in the plane of two observable quantities, with $N({\rm OH^+})/N({\rm H_2O^+})$ plotted on the horizontal axis and $N({\rm OH^+})/N({\rm H})$ on the vertical axis.  Once again, overplotted contours show the predictions of our diffuse cloud model, with contours of visual extinction shown in blue, and contours of CRIR shown in red.  Blue contours are labeled with $\zeta_p({\rm H})/n_{50}$ in units of $10^{-16}\,\rm s^{-1}$, and red contours with $A_{\rm V}({\rm tot})$ in mag.  All the model predictions apply to a UV radiation field with $\chi_{UV}/n_{50} = 1.$  Figure 8, like Figure 4, shows the corresponding predictions obtained from an analytic treatment, in this case that of Neufeld et al.\ (2010) and I15 (their equations 12 and 15).  Here, several simplifying assumptions were made: (1) a constant fraction, $\epsilon=0.07$, of H ionizations lead to OH$^+$, with the value of $\epsilon$ ``calibrated" by observations of a single source toward which H$_3^+$, OH$^+$, and $\rm H_2O^+$ are all observed (Indriolo et al.\ 2012); (2) $\rm H_2O^+$ is formed exclusively by reaction of OH$^+$ with H$_2$; (3) OH$^+$ and H$_2$O$^+$ are destroyed exclusively by dissociative recombination and reaction with H$_2$; (4) the electron abundance is equal to the gas-phase carbon abundance; and (5) the H$_2$ fraction in a given cloud is constant (red contours) throughout the zone in which OH$^+$ and $\rm H_2O^+$ are present.  A comparison of Figures 7 and 8 indicates that the simple analytic treatment adopted in I15 significantly overestimates the $N({\rm OH^+})/N({\rm H})$ ratio when the CRIR is large.  This behavior results from a breakdown of assumption (4) above; for large CRIRs, H ionization contributes significantly to the electron abundance, thereby increasing the OH$^+$ destruction rate.  Finally, in Figure 9, we have transformed the coordinate system so that contours of observable quantities [$N({\rm OH^+})/N({\rm H_2O^+})$ and $N({\rm OH^+})/N({\rm H})$] are plotted in the plane of model parameters [$\zeta_p({\rm H})/n_{50}$ and $A_{\rm V}({\rm tot})$].

\begin{figure}
\includegraphics[width=14 cm]{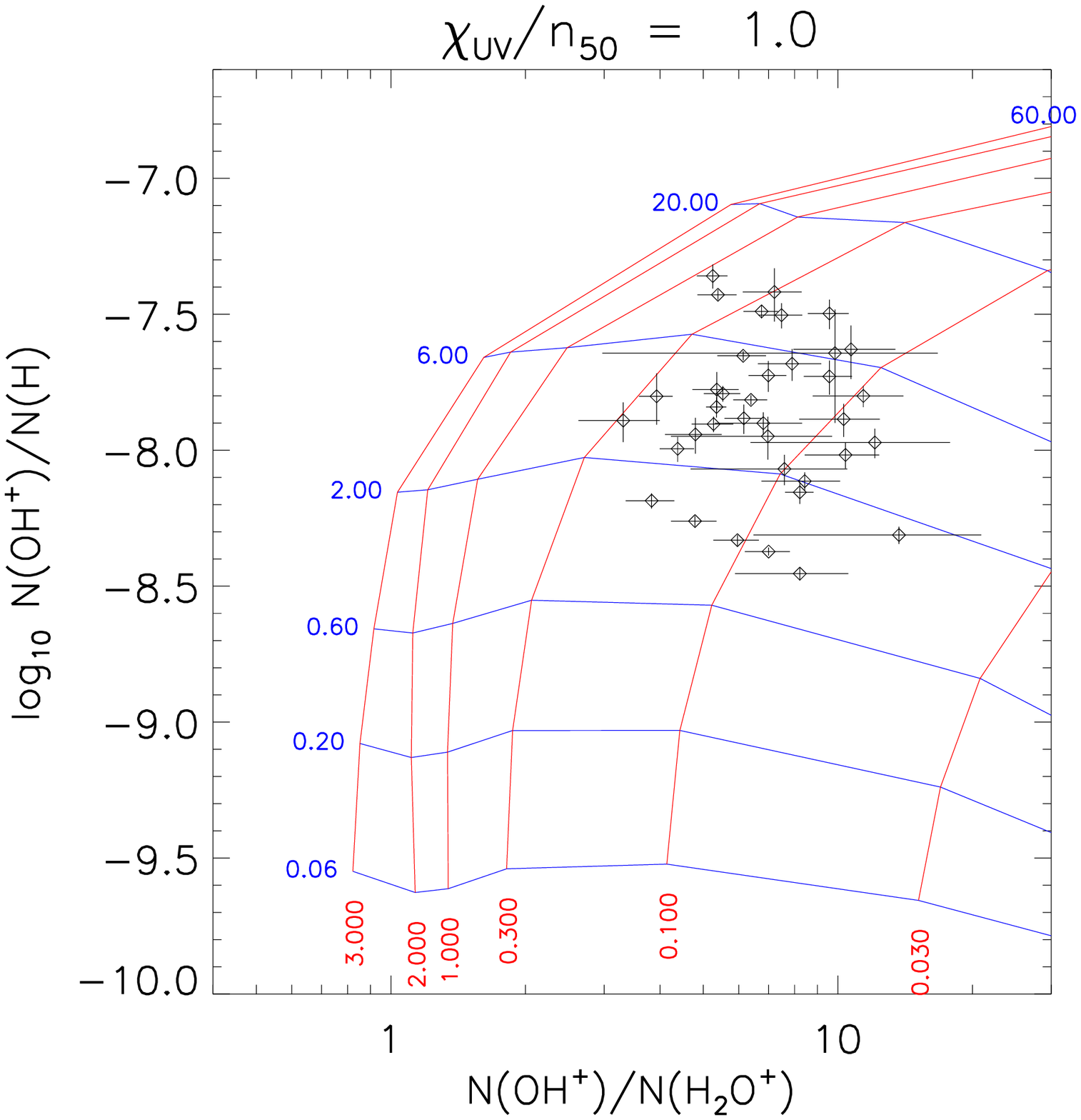}
\caption{
${N({\rm OH^+}) / N({\rm H_2O^+})}$ and ${N({\rm OH^+}) / N({\rm H})}$ column density 
ratios predicted for
diffuse and translucent molecular clouds with $\chi_{\rm UV}/n_{50} = 1$, 
where $\chi_{\rm UV}$ is the incident
radiation field in Draine (1978) units and $n_{\rm H} = 50\,n_{50}\,\rm cm^{-3}$
is the density of H
nuclei. Results are shown in the plane of ${N({\rm OH^+}) / N({\rm H_2O^+})}$
and ${N({\rm OH^+}) / N({\rm H})}$,
with contours of
visual extinction, $A_{\rm V}({\rm tot})$, 
shown in red and contours of $\zeta_p({\rm H})/n_{50}$ shown in blue
(where $\zeta_p({\rm H}) \sim \zeta_t({\rm H}_2)/2.3$ is the primary cosmic-ray ionization rate per H nucleus and $\zeta_t({\rm H}_2)$ is the total cosmic-ray ionization rate per H$_2$ molecule.)
}
\end{figure}

\begin{figure}
\includegraphics[width=14 cm]{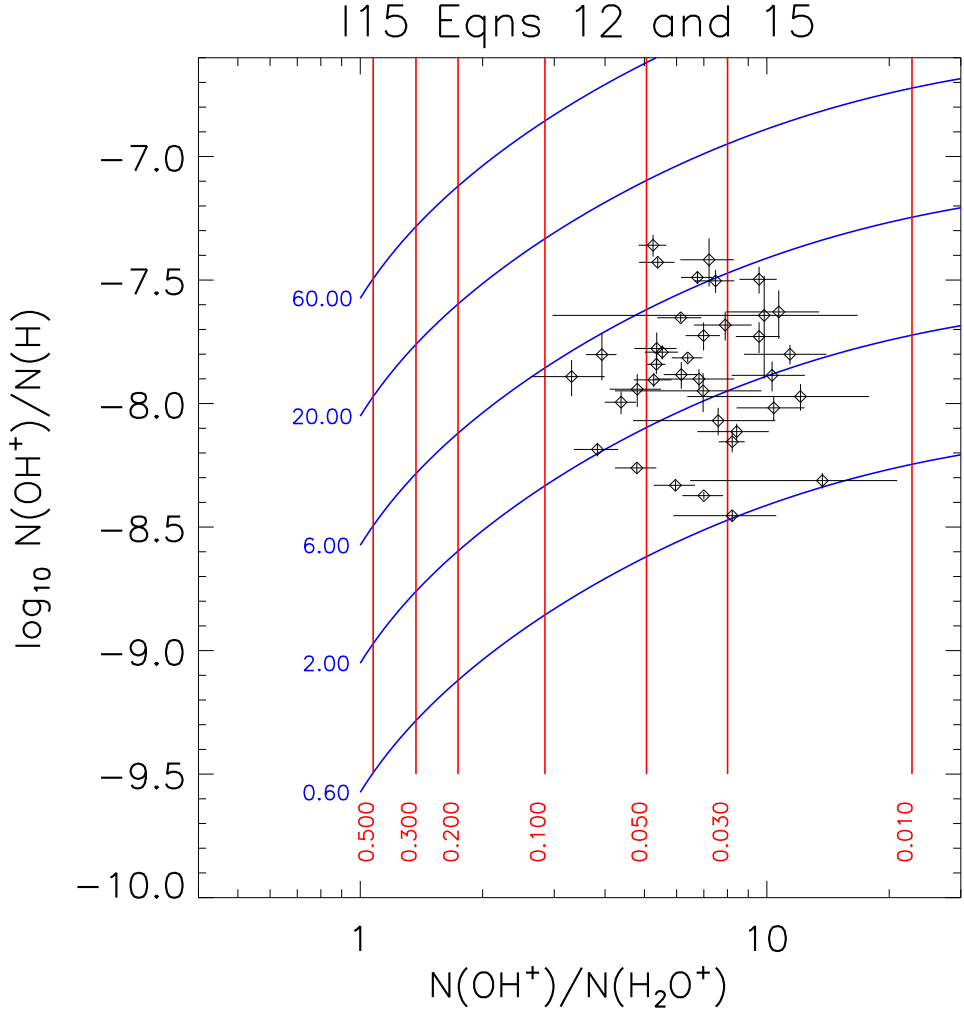}
\caption{
Same as Figure 7, but with the column density ratio
computed using the simple analytic approximations adopted by N15, and with the red
contours being contours of molecular fraction.
}
\end{figure}

\begin{figure}
\includegraphics[width=14 cm]{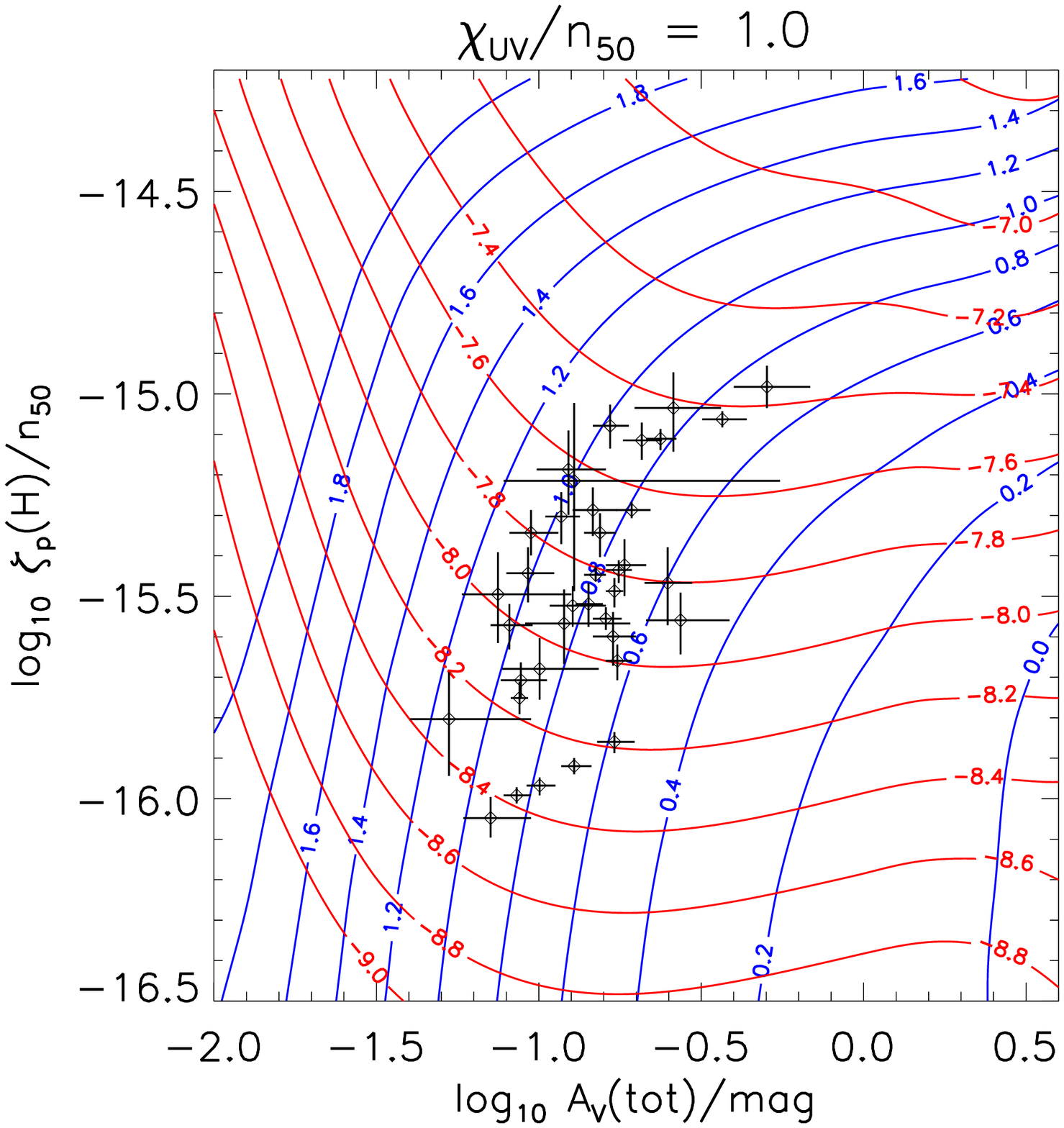}
\caption{
Same as Figure 7, but now with contours of
${N({\rm OH^+}) / N({\rm H})}$ and ${N({\rm OH^+}) / N({\rm H_2O^+})}$  in the plane of the model parameters 
$A_{\rm V}({\rm tot})$ and $\zeta_p({\rm H})/n_{50}$.
Blue contours are labeled with ${\rm log}_{10}[N({\rm OH^+}) / N({\rm H_2O^+})]$, and red contours are labeled with ${\rm log}_{10}[N({\rm OH^+}) / N({\rm H})]$}
\end{figure}

For the diffuse atomic ISM, I15 have presented observations of $N({\rm OH}^+)$ and $N({\rm H_2O^+})$ along 20 Galactic sight-lines toward background sources of bright submillimeter continuum emission.  Combined with HI 21 cm observations obtained by Winkel et al.\ (2017), these observations permit reliable absorption-line determinations of $N({\rm OH^+})/N({\rm H_2O^+})$ and $N({\rm OH^+})/N({\rm H})$ in 37 distinct velocity intervals arising in foreground diffuse atomic gas within the Galactic disk.  For 15 of these velocity intervals, observations of ArH$^+$ absorption are also available (S14).   The measured values of  $N({\rm OH^+})/N({\rm H_2O^+})$ and $N({\rm OH^+})/N({\rm H})$ are represented by diamonds  in Figures 7 -- 8, along with their 1 $\sigma$ error bars, and the corresponding clouds parameters are shown in Figure 9, along with their 68\% confidence intervals.  

One simplifying assumption adopted here is that the $N({\rm OH}^+)$ and $N({\rm H_2O^+})$ absorptions originate in the same gas as the HI 21 cm absorption.  However, as shown in Paper I, an analysis of the $\rm OH^+$, $\rm H_2O^+$ and $\rm ArH^+$ column densities shows that a single population of clouds cannot account simultaneously for the observations.  Instead, the measured column densities require at least two distinct populations of diffuse atomic clouds: (1) a population of smaller clouds, which are primarily responsible for the observed ArH$^+$ absorption, with a total visual extinction of at most 0.02 mag per cloud and a column-averaged molecular fraction in the range $10^{-5}$ to 10$^{-2}$; and (2) a population of somewhat larger clouds, primarily responsible for the observed OH$^+$ and H$_2$O$^+$ absorption, in which the column-averaged molecular fraction is $\sim 0.2$.  Because part of the observed 21 cm absorption originates in population (1) above, the $N({\rm OH^+})/N({\rm H})$ ratio in population (2) can be larger than the measured ratio.  This effect will be discussed further in \S 4.1 below.

\section{Discussion}

\subsection{Comparison with previous estimates of the CRIR in the diffuse ISM}

\subsubsection{Estimates obtained from observations of molecular ions}

In Figure 10, we present a comparison of the CRIRs derived in the present study with previous estimates obtained by N12 and I15.  
The top row in Figure 10 shows the CRIRs, $\zeta_p({\rm H})/n_{50}$, 
derived for diffuse atomic gas in 
which the column densities of $\rm OH^+$, $\rm H_2O^+$, $\rm ArH^+$ and H have all been measured.  Here, blue diamonds represent the values obtained previously using a simple analytic treatment of the chemistry, whereas red and magenta symbols indicate the results obtained using the detailed diffuse clouds models.  The CRIRs indicated by the red diamonds were computed without any correction for the presence of small ArH$^+$-containing clouds that may contribute significantly to the HI column density but not the OH$^+$ and H$_2$O$^+$ column densities.     The results shown by the magenta diamonds apply the necessary correction, using the methodology described in Paper I.  Here, a simultaneous fit to the $N({\rm OH}^+)/N({\rm H})$, $N({\rm H_2O}^+)/N({\rm H})$, and $N({\rm ArH^+})/N({\rm H})$ ratios was obtained for a combination of the smaller and larger cloud types described in \S 3.2 above.  In this analysis, we assumed the standard UV radiation field, $\chi_{\rm UV}/n_{50}=1$, and varied four free parameters: $\zeta_p({\rm H})/n_{50}$ (assumed to be the same in both cloud types); the fraction of atomic H in the population of smaller clouds, $f_{\rm S}$;  the total visual extinction across an individual small cloud, $A_{\rm V}({\rm tot})_{\rm S}$; and the total visual extinction across an individual large cloud, $A_{\rm V}({\rm tot})_{\rm L}$.  Because there are only three observables -- $N({\rm OH}^+)/N({\rm H})$, $N({\rm H_2O}^+)/N({\rm H})$, and $N({\rm ArH^+})/N({\rm H})$ -- the problem is underconstrained and thus there is a range of CRIRs that can satisfactorily match the data for any given velocity interval.   This range is reflected in the error bars on the magenta diamonds.  In deriving the range of acceptable values for the CRIR, we apply the additional constraint that $f_{\rm S} \le 0.5,$ i.e. that the population of larger clouds contains at least one-half of the observed HI.  This constraint is motivated by the observational fact that the HI absorption spectra are typically more similar to the OH$^+$ and $\rm H_2O^+$ spectra than they are to the ArH$^+$ spectra (Neufeld et al.\ 2015).  The labels underneath the plotted points indicate the background source for each CRIR determination and the velocity interval to which the determination applies (in km~s$^{-1}$ with respect to the local standard of rest.)

\begin{figure}
\includegraphics[width=10 cm]{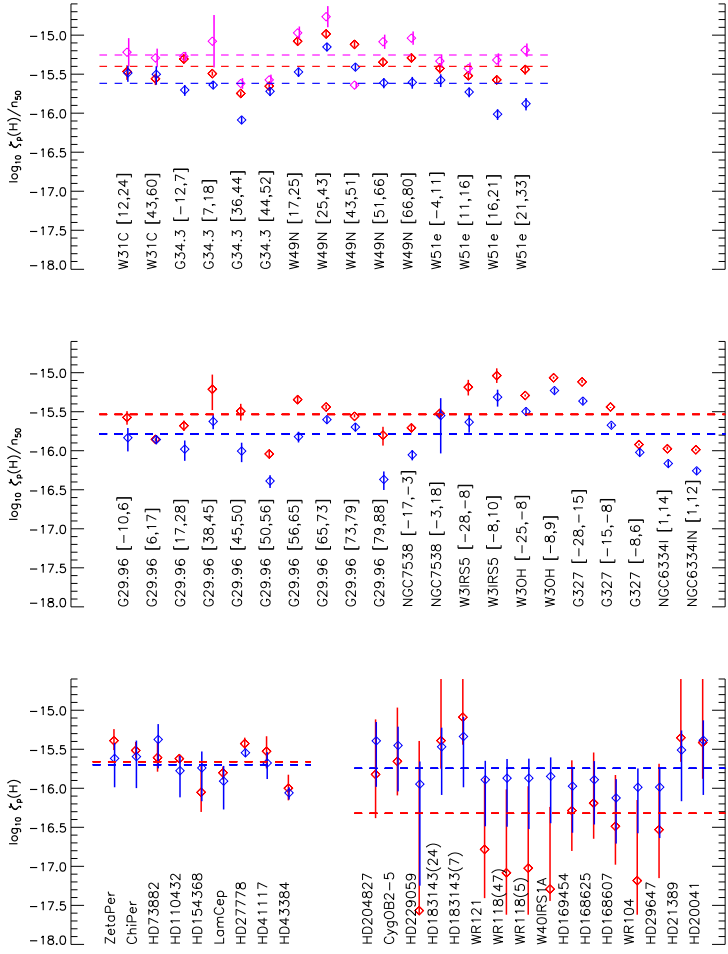}
\caption{Comparison of the CRIRs derived in the present study with previous estimates obtained by N12 and I15.  The labels underneath the plotted points indicate the background source for each CRIR determination and the velocity interval to which the determination applies.  Top row: $\zeta_p({\rm H})/n_{50}$, 
derived for diffuse atomic gas in 
which the column densities of $\rm OH^+$, $\rm H_2O^+$, $\rm ArH^+$ and H have all been measured.  Middle row: $\zeta_p({\rm H})/n_{50}$, derived for diffuse atomic gas in which ArH$^+$ observations are not available.  Bottom row: CRIRs derived for diffuse molecular gas from observations of H$_3^+$.  Bottom left: clouds in which H$_2$ is measured directly.  Bottom right: clouds in which H$_2$ is only measured indirectly. Blue diamonds: values obtained previously using a simple analytic treatment of the chemistry (IM12 or I15).   Red diamonds: values obtained using detailed diffuse cloud models, but without any correction for HI in the small diffuse atomic clouds responsible for the observed ArH$^+$ absorption.  Magenta diamonds: values obtained using detailed diffuse cloud models with the inclusion of the aforementioned correction.}
\end{figure}

The middle row of Figure 10 shows the CRIRs, $\zeta_p({\rm H})/n_{50}$, derived for diffuse atomic gas in which ArH$^+$ observations are not available.  Here, only the blue and red diamonds can be presented.  

Finally, the bottom row of Figure 10 shows the CRIRs derived for diffuse molecular gas from observations of H$_3^+$.  The nine determinations on the left are the most reliable, because they apply to clouds in which H$_2$ is measured directly, while the remaining determinations -- with correspondingly larger uncertainties -- apply to clouds where only indirect H$_2$ measurements are available.  For the bottom row of Figure 10, we have used the density estimates adopted by I15 to present values for $\zeta_p({\rm H})$ itself, rather than $\zeta_p({\rm H})/n_{\rm H}$.

For each set of CRIR determinations plotted in Figure 10, the mean values are indicated by dotted horizontal lines with the same color-coding as the diamonds.  For diffuse atomic clouds (top and middle rows), our detailed cloud models yield CRIR-estimates that are systematically larger than those obtained with the simplifying assumptions used in previous studies, by an average factor of 1.7 (0.23 dex).   Applying a correction for HI in small diffuse clouds, which can be implemented in gas where ArH$^+$ is observed (top row), increases our estimates of the CRIR by a further factor of 1.4 (0.15 dex).  

For diffuse molecular clouds probed by H$_3^+$ and direct measurements of H$_2$ (bottom left), the CRIRs derived from the detailed cloud models are in excellent agreement with the simple analytical treatment, with the average discrepancy being only $9\%$ (0.04 dex).  For those diffuse molecular clouds without direct measurements of H$_2$, however, the CRIRs derived from the detailed models are, on average, a factor 3.8 (0.58 dex) lower than those derived using the simple analytic estimates.  For these clouds, which are of larger $A_{\rm V}({\rm tot})$ than any cloud in which H$_2$ can be measured directly by UV absorption line observations, the electron abundance falls below the gas-phase carbon abundance; as a result, the H$_3^+$ destruction rate is overestimated in the simple analytic treatment of IM12, leading to an overestimate of the CRIR required to fit a given value of $N({\rm H}_3^+).$ 

\subsubsection{Estimates obtained from radio recombination line observations of atomic ions}

As originally discussed by Shaver (1976) and Sorochenko \& Smirnov (1987; hereafter SS87), radio recombination lines (RRLs) from atomic ions provide an alternate probe of the CRIR in the diffuse neutral ISM.   Here, H$^+$ is produced by cosmic-ray ionization, whereas C$^+$ is produced by photoionization.  Thus the strength of hydrogen radio recombination lines (HRRLs) relative to that of carbon radio recombination lines (CRRLs) is an increasing function of $\zeta_p({\rm H})/n_{\rm H}$.  

To date, the sight-line to the Cas A supernova remnant is the best-studied case in which RRLs have been observed from the diffuse ISM.  Oonk et al.\ (2017; hereafter O17) 
have recently reported new measurements of RRL strengths for this sight-line, derived from high-quality interferometric data obtained from the Low Frequency Array (LOFAR) and the Westerbork Synthesis Radio Telescope (WSRT).  These included WSRT detections of H$n\alpha$ line emission -- with principal quantum number $n$ in the range 257 to 278 -- 
from a cold cloud located in the Perseus arm at a velocity of $-47\,\rm km \,s^{-1}$ relative to the Local Standard of Rest (LSR).  
With the use of a new model (Salgado et al.\ 2016) for the level populations of Rydberg states of C, together with an analysis of the observed line widths, O17 derived an CRRL emission measure for the 
$-47\,\rm km \,s^{-1}$ component of
$$EM_{\rm C} = \int n({\rm C}^+) n_e dz = 0.056 \pm 0.014 \rm \, cm^{-6} \rm pc , \eqno(6)$$ an 
electron temperature of $85 \pm 5$~K, and an electron density of $0.040 \pm 0.05  \rm \,cm^{-3}$. 
O17's preferred model for the $-47\,\rm km \,s^{-1}$ component is a diffuse molecular cloud with a sheet-like geometry, observed at an oblique inclination, and with a density $n_{\rm H} \sim 2.9 \times 10^2 \rm cm^{-3}$ and line-of-sight column density $N_{\rm H} \sim 3
\times 10^{22} \rm cm^{-3}$.

\begin{figure}
\includegraphics[width=15 cm]{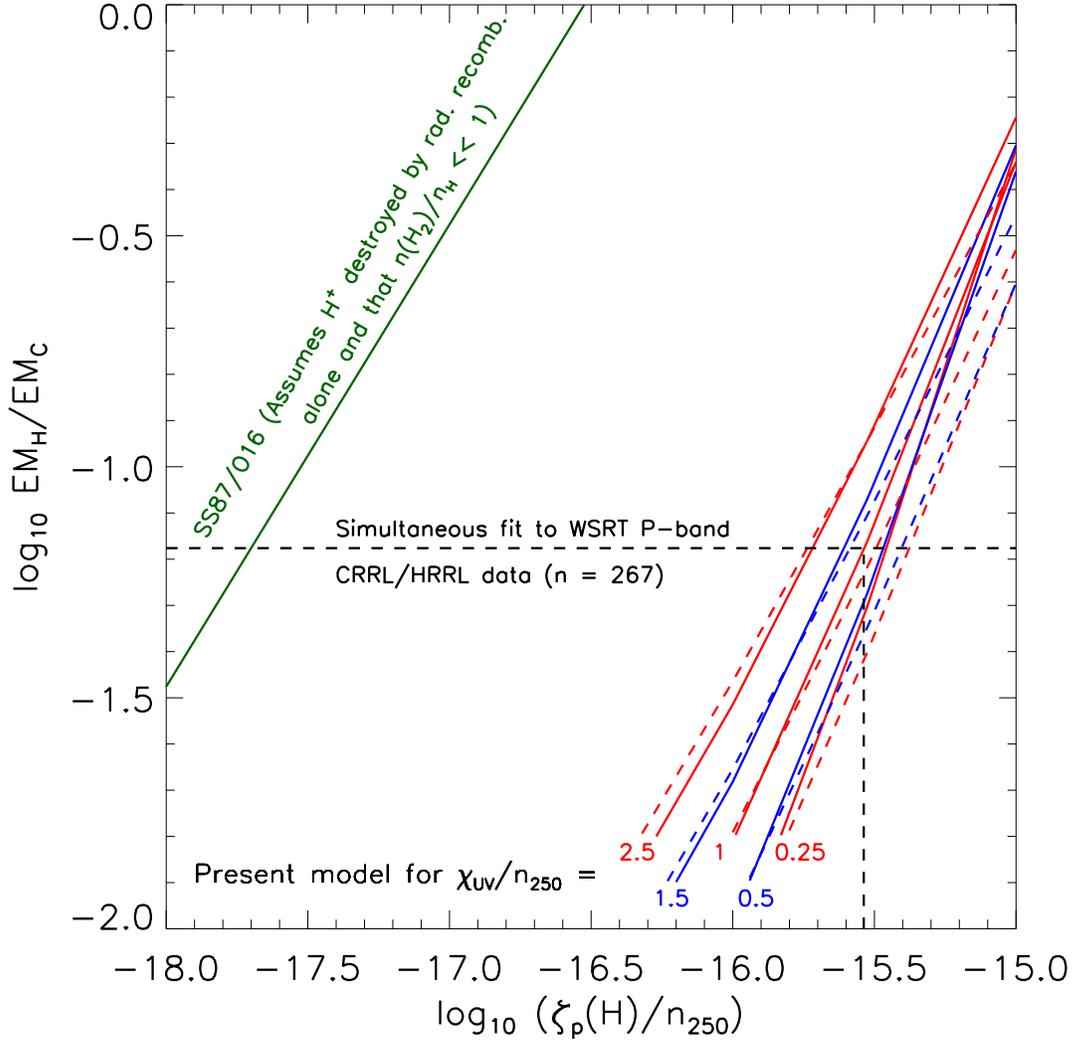}
\caption{Ratio of emission measures, $EM_{\rm H}/EM_{\rm C}$, predicted by our diffuse clouds models as a function of the CRIR.  Results are shown for five values of the UV radiation field, $\chi_{\rm UV} /n_{250}$ = 0.25, 0.5, 1, 1,5, and 2.5; and for two values of the cloud extinction, $A_V({\rm tot})$ = 1 mag (dashed curves) and 5 mag (solid curves). Green curve: $EM_{\rm H}/EM_{\rm C}$ predictions obtained using the simple SS87 analysis (see text).}
\end{figure}

Given these values for electron temperature and density, along with the observed strengths of the HRRLs detected with WSRT, O17 derived an HRRL emission measure, $EM_{\rm H} = 0.0036 \rm \,cm^{-6}\,\rm pc$, implying $EM_{\rm H}/EM_{\rm C}=0.064.$  Using a simple analysis due to SS87, in which radiative recombination is assumed to dominate the destruction of H$^+$ and HI is assumed to be the dominant reservoir of H nuclei, O17 found that a total CRIR of $\zeta_t({\rm H}) = 2.5 \times 10^{-18}\rm \,s^{-1}$ was needed to match the observed $EM_{\rm H}/EM_{\rm C}$ ratio, a CRIR value much smaller than those derived from observations of molecular ions.  O17 noted, however, that a much larger CRIR could be required (e.g.\ Liszt 2003) if the neutralization of H$^+$ by small grains enhances the destruction of H$^+$; the above value is therefore a strict lower limit.  

In Figure 11, we plot the values of $EM_{\rm H}/EM_{\rm C}$ predicted by our diffuse clouds model as a function of the CRIR.  Results are shown for five values of the UV radiation field, $\chi_{\rm UV} /n_{250}$ = 0.25, 0.5, 1.0, 1.5, and 2.5; and for two values of the cloud extinction, $A_V({\rm tot})$ = 1 mag (dashed) and 5 mag (solid).  The $EM_{\rm H}/EM_{\rm C}$ ratio is an increasing function of $\chi_{\rm UV} /n_{250}$, because larger UV radiation fields enlarge the region within which H is primarily atomic.  Moreover, $EM_{\rm H}/EM_{\rm C}$ is roughly independent of $A_V({\rm tot})$ for values typical of diffuse molecular clouds, because both the CRRL and HRRL emissions occur relatively close to cloud surfaces.  For comparison, the green line shows the much larger $EM_{\rm H}/EM_{\rm C}$ values predicted using the simple SS87 analysis.  In addition to the destruction of H$^+$ in reactions with neutral or negatively-charged PAHs, two further effects reduce the $EM_{\rm H}/EM_{\rm C}$ ratio below the SS87 predictions.  First -- as discussed, for example, by Sorochenko \& Smirnov (2010) -- the H$_2$ fraction becomes significant within the CRRL-emitting region, so that the atomic hydrogen abundance and HRRL line emission are diminished accordingly.  Second, charge transfer reactions of atomic oxygen dominate radiative recombination as a destruction process for H$^+$, so that even in the absence of PAH-assisted recombination the H$^+$ abundance would be reduced sharply.

The horizontal dashed line in Figure 11 shows the $EM_{\rm H}/EM_{\rm C}$ ratio obtained by O17 from a simultaneous fit to the CRRL and HRRL strengths measured for the $-47\,\rm km \,s^{-1}$ cloud using WSRT.  With inclusion of the various factors that reduce the $EM_{\rm H}/EM_{\rm C}$ ratio far below the predictions of SS87, and given the density estimate of O17, 
our best-fit CRIR is $\zeta_p({\rm H}) = 2.9 \times 10^{-16}\rm \,s^{-1}$, for an assumed $\chi_{\rm UV}$ of 1.  This value is roughly two orders of magnitude larger that the lower limit obtained by O17,  $\zeta_t({\rm H}) = 1.5\, \zeta_p({\rm H})  = 2.5 \times 10^{-18}\rm \,s^{-1}$, and is very typical of the estimates derived from molecular abundances.

One caveat applies to discussion given above.  In addition to detecting Hn$\alpha$ RRLs in the range $n=257$ to 278 using WSRT, O17 also obtained upper limits on the strengths of lower frequency H$n\alpha$ RRLs (with $n \sim 500$) using LOFAR.  To reconcile the HRRL detections obtained with WSRT with these upper limits from LOFAR, O17 required a somewhat larger assumed electron density ($\sim 0.065-0.11\,\rm cm^{-3}$) and somewhat lower assumed gas temperature ($\sim 30 - 50$~K) than those derived from their analysis of the CRRL.  Properly investigatng this discrepancy will require calculations that are beyond the scope of the present study.  Instead of simply computing the relative emission measures for C$^+$ and H$^+$, future models will need to integrate the individual RRL strengths over the cloud, taking account of the varying temperature and electron density to obtain predictions for the H$n\alpha$ and C$n\alpha$ line strengths as a function of $n$.

\subsection{Mean and dispersion of the CRIR}

With the aid of the detailed diffuse cloud models described in this paper, we obtain the estimates of the CRIR given in Table 2.  Results are given here for 4 subsets of the data.  From left to right, these are (1) diffuse molecular clouds in which $\rm H_3^+$ and H$_2$ are measured directly and gas density estimates are available from observations of C$_2$;
(2) diffuse atomic gas in which OH$^+$, H$_2$O$^+$ and HI have been measured but not ArH$^+$; (3) diffuse atomic gas in which OH$^+$, H$_2$O$^+$, ArH$^+$ and HI have all been measured; (4) all diffuse atomic gas [i.e. the union of subsets (2) and (3)].  In obtaining our estimate of the average CRIR in subset (4), we have applied the mean correction for HI in small clouds obtained in subset (3) to subset (2) in which ArH$^+$ measurements are unavailable.

For each of these subsets, we list the sample size, the mean of ${\rm log}_{10} \zeta_p({\rm H})$ or ${\rm log}_{10}[\zeta_p({\rm H})/n_{50}]$ and its standard error, the corresponding values of $\zeta_p({\rm H})$ or $\zeta_p({\rm H})/n_{50}$, and the dispersion, $\sigma_{\rm BE}$, of the best estimates of ${\rm log}_{10} \zeta_p({\rm H})$ or ${\rm log}_{10}[\zeta_p({\rm H})/n_{50}]$ plotted in Figure 10.  Because our estimates of these quantities have known uncertainties, 
resulting from uncertainties in the column-density measurements upon which they are based, $\sigma_{\rm BE}$ is   
an upper limit on the true dispersion of ${\rm log}_{10} \zeta_p({\rm H})$ or ${\rm log}_{10}[\zeta_p({\rm H})/n_{50}]$.  On the assumption that the errors in these quantities are Gaussian, and that the actual distribution of $\zeta_p({\rm H})$ or ${\rm log}_{10}[\zeta_p({\rm H})/n_{50}]$ is log normal, we may estimate the true dispersion, $\sigma_{\rm T}$, of either quantity from the equation 
$$\chi^2_{red} = {1 \over N-1}\sum_i \bigl[ (x_i - x_m)^2/(\sigma_{\rm T}^2 + \sigma_{i}^2)\bigr] = 1, \eqno(5)$$ where $N$ is the number of objects in the sample, $x_i$ is the best estimate of ${\rm log}_{10} \zeta_p({\rm H})$ or ${\rm log}_{10}[\zeta_p({\rm H})/n_{50}]$ for the {\it i}th object, $\sigma_i$ is the uncertainty in $x_i$, and
$x_m$ is the mean of the $x_i$.  
  Values of $\sigma_{\rm T}$ are given in Table 2.

Entries in boldface represent the key results obtained from the present study.  For diffuse molecular gas probed by H$_3^+$, we obtain $-15.63 \pm 0.09$ (standard error) for the mean of ${\rm log}_{10} \zeta_p({\rm H})$.  This value is a factor 1.2 times as large as the previous estimate presented by IM12.
We estimate the true dispersion of ${\rm log}_{10} \zeta_p({\rm H})$ as 0.09.  For diffuse atomic clouds clouds probed by OH$^+$, $\rm H_2O^+$ and ArH$^+$,  we obtain $-15.34 \pm 0.05$ (standard error) for ${\rm log}_{10}[\zeta_p({\rm H})/n_{50}]$.   This value is a factor 2.6 times as large as the previous estimate presented by I15.  We estimate the true dispersion of ${\rm log}_{10}[\zeta_p({\rm H})/n_{50}]$ as 0.23.

One striking feature of our CRIR estimates is their remarkably low dispersions.  In the case of the diffuse molecular clouds, the value of 0.09 for $\sigma_{\rm T}$ corresponds to a cloud-to-cloud variation of only $\sim 25\%$.  Moreover, our analysis did not include uncertainties in the density estimates derived from C$_2$ observations, so the value of 0.09 is really an upper limit.  In the case of the diffuse atomic clouds, the value of 0.23 for $\sigma_{\rm T}$ 
corresponds to a variation by only a factor of 1.7.  In this case, the dispersion of ${\rm log}_{10}[\zeta_p({\rm H})/n_{50}]$ includes both intrinsic variations in the CRIR and intrinsic variations in the density.  Here again, 0.23 dex is strict upper limit on any variations in the CRIR.   

One important caveat should be noted in the case of diffuse molecular clouds: the mean and dispersion of the CRIR given above applies specifically to a sample of stars towards which H$_3^+$ (and H$_2$) {\it have been detected}.  The set of sight-lines discussed by IM12 also include multiple cases with H$_3^+$ {\it non-detections}, and in some of these the upper limits inferred by IM12 for the CRIR are significantly smaller than the average value.  In particular, IM12 noted that the nearby Ophiuchus-Scorpius region appears to exhibit an abnormally low CRIR.

\begin{deluxetable}{lcccc}
\tabletypesize{\scriptsize}
\tablewidth{0pt}
\tablecaption{Estimates of the CRIR: mean values and dispersions}
\tablehead{Method & H$_3^+$ & OH$^+$ and $\rm H_2O^+$ & OH$^+$ and $\rm H_2O^+$ & OH$^+$ and $\rm H_2O^+$ \\
& & without ArH$^+$ & with ArH$^+$ & (all)}
\startdata
Sample size & 7 & 22 & 15 & 37\\
\\
$< {\rm log}_{10} \zeta_p({\rm H}) > $ (Present work) & $\bf -15.63 \pm 0.09$ \\
$< {\rm log}_{10} \zeta_p({\rm H}) > $ (IM12) & $-15.73$ \\
$10^{< {\rm log}_{10} \zeta_p({\rm H}) >} /10^{-16}\,\rm s^{-1}$ (Present work) & $2.3 \pm 0.6$ \\
$10^{< {\rm log}_{10} \zeta_p({\rm H}) >} /10^{-16}\,\rm s^{-1}$ (IM12) & 1.9 \\
$\sigma_{\rm BE}[{\rm log}_{10} \zeta_p({\rm H})]$ (Present work) & 0.23 \\
$\sigma_{\rm BE}[{\rm log}_{10} \zeta_p({\rm H})]$ (IM12) & 0.24 \\
$\sigma_{\rm T}[{\rm log}_{10} \zeta_p({\rm H})]$ (Present work) & {\bf 0.09} \\
\\
$< {\rm log}_{10}[\zeta_p({\rm H})/n_{50}] > $ (Present work) && $-15.54 \pm 0.07$ & $-15.25 \pm 0.07$ & $\bf -15.34 \pm 0.05$ \\
$< {\rm log}_{10}[\zeta_p({\rm H})/n_{50}] > $ (I15) &&&& --15.77\\
$10^{< {\rm log}_{10} [\zeta_p({\rm H})/n_{50}] >}/10^{-16}\,\rm s^{-1} $ (Present work) &&  $2.9 \pm 0.5$ & $5.6 \pm 0.9$ & $\bf 4.6 \pm 0.5$ \\
$10^{< {\rm log}_{10} [\zeta_p({\rm H})/n_{50}] >} /10^{-16}\,\rm s^{-1} $ (I15) &&&& 1.8 \\
$\sigma_{\rm BE}[{\rm log}_{10}[\zeta_p({\rm H})/n_{50}]]$ (Present work) && 0.31 & 0.25 & 0.28 \\
$\sigma_{\rm T}[{\rm log}_{10}[\zeta_p({\rm H})/n_{50}]]$ (Present work) && 0.26 & {\bf 0.23}  \\
\\
Notes & (a) & (b) & (c) & (d) \\
\enddata
\tablenotetext{a}{Includes only sight-lines for which H$_2$ has been measured directly and H$_3^+$ has been detected, and assumes that $\chi_{UV}/n_{250} = 1$}
\tablenotetext{b}{Assumes that 100 percent of the observed HI is present within the larger clouds responsible for OH$^+$ ,and $\rm H_2O^+$, \bk{and that $\chi_{UV}/n_{50} = 1$}}
\tablenotetext{c}{Assumes that up to 50 percent of observed HI may be present within the smaller clouds responsible for ArH$^+$, \bk{and that $\chi_{UV}/n_{50} = 1$}}
\tablenotetext{d}{In cases where ArH$^+$ is not observed, a correction for HI in smaller clouds is applied, using the average correction factor obtained when $N({\rm ArH}^+)$ is measured. A $\chi_{UV}/n_{50}$ value of 1 is assumed.}

\end{deluxetable}

\begin{figure}
\includegraphics[width=15 cm]{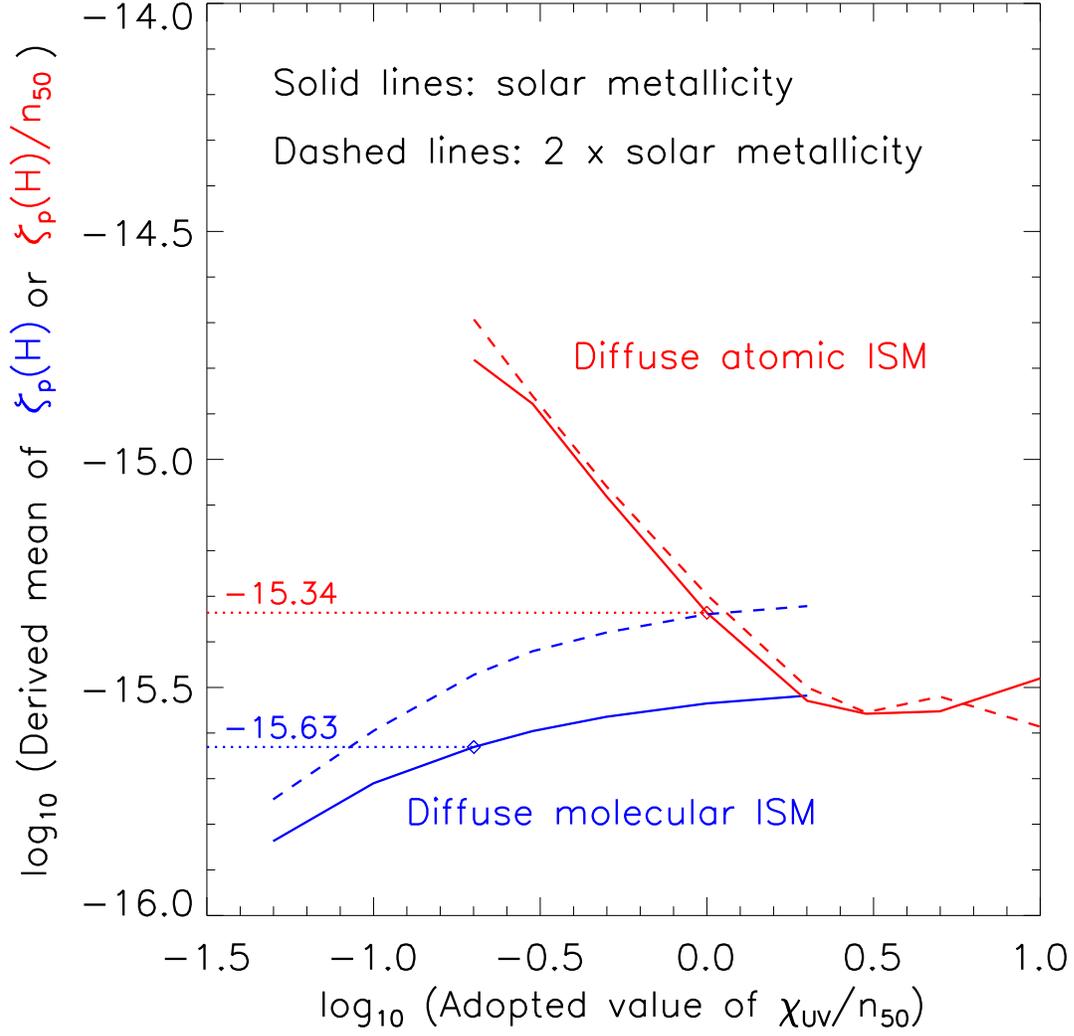}
\caption{Dependence of the mean derived CRIRs upon the assumed values of $\chi_{\rm UV}/n_{\rm 50}$ for $Z=Z_{\rm std}$ (solid curves) and $Z=2\,Z_{\rm std}$ (dashed curves).  Blue curves: mean CRIRs derived for diffuse molecular clouds (i.e.\ from H$_3^+$).  Red curves: mean CRIRs derived for diffuse atomic clouds}
\end{figure}

\subsection{Dependence of our CRIR estimates on the assumed UV radiation field and gas metallicity}

All the results presented in Figures 1 -- 10 were obtained under the assumptions that $\chi_{\rm UV}/n_{\rm 50} =1$ in diffuse atomic clouds, $\chi_{\rm UV}/n_{\rm 50} =0.2$ in diffuse molecular clouds, and that the abundances of the heavy elements have the standard values adopted for the gas-phase in diffuse interstellar material in the solar neighborhood, $Z=Z_{\rm std}$.  To examine the dependence of the mean CRIRs we derive (Table 2) upon these assumptions, we have repeated the entire analysis described in \S4.2 for a range of assumed $\chi_{\rm UV}/n_{\rm 50}$ and for two assumed metallicities, $Z=Z_{\rm std}$ and $Z=2\,Z_{\rm std}$.  In Figure 12, we show the dependence of the mean derived CRIRs upon the assumed values of $\chi_{\rm UV}/n_{\rm 50}$ for $Z=Z_{\rm std}$ (solid curves) and $Z=2\,Z_{\rm std}$ (dashed curves).  The blue curves show the mean CRIRs derived for diffuse molecular clouds (i.e.\ from H$_3^+$), and the red curves show those derived for diffuse atomic clouds (i.e.\ from OH$^+$, $\rm H_2O^+$, and ArH$^+$).  The former show that the CRIR derived from H$_3^+$ observations are roughly proportional to $Z/Z_{\rm std}$; this behavior results because the abundance of electrons, which are primarily responsible for the destruction of H$_3^+$ given the typical CRIRs in the Galactic disk, is roughly equal to the gas-phase carbon abundance.
The red curves show that the CRIR derived from observations of OH$^+$, $\rm H_2O^+$, and ArH$^+$ are almost independent of $Z$ for the two cases we examined; here, the increased abundance of electrons is roughly balanced by the increased abundances of gas-phase O and Ar.  The dependences of the mean derived CRIRs on the assumed value of $\chi_{\rm UV}/n_{\rm 50}$ reflects a complex interplay of factors: in the case of the mean CRIR in diffuse molecular clouds, the final dependence is quite weak, whereas the mean CRIR in diffuse atomic clouds decreases roughly as $(\chi_{\rm UV}/n_{\rm 50})^{-0.7}$ for the typical values in the Galactic disk.  Also shown on Figure 12 are the mean values presented in Table 2, which correspond to $\chi_{\rm UV}/n_{\rm 50}=0.2$ -- or equivalently $\chi_{\rm UV}/n_{\rm 250}=1$ -- in diffuse molecular clouds, and $\chi_{\rm UV}/n_{\rm 50}=1$ in diffuse atomic clouds.

\subsection{Sensitivity to the assumed reaction rates}

The uncertainty estimates presented in this paper are statistical in nature and do not include systematic uncertainties inherent in the diffuse cloud models.  Because the model is based upon a large number of assumptions about fundamental physical and chemical processes -- including the rate coefficients for a large number of chemical reactions -- a quantitative analysis of the systematic uncertainies is impractical.  We can, however, identify several key reactions with rate coefficients that are important in determining what CRIR is needed to match the available data.  These are listed in Table 3, along with the rate coefficients we adopted and the primary bibliographic references of relevance.  In all cases, the values adopted are the same as those in H12\footnote{We note here a typographic error in H12 Table 1, where the temperature dependence for reaction $k_7$ has a sign error in the exponent.  (This error affected only Table 1, because the correct exponent was used in all the calculations performed by H12).}  Based upon the analytic treatment of $\rm OH^+$ and $\rm H_2O^+$ given by H12 (their Appendix B), the CRIR is expected to show the following dependences:

\begin{deluxetable}{lll}
\tablewidth{0pt}
\tablecaption{Key reaction rates}
\tablehead{Reaction & Adopted rate coefficient ($\rm cm^3 s^{-1}$) & Primary reference}
\startdata
$\rm H_3^+ + e \rightarrow products$ & $k_1=6.8 \times 10^{-8} (T/{\rm 300\, K})^{-0.5}$ & McCall et al.\ (2003)\\
$\rm H^+ + PAH \rightarrow H + PAH^+$ & $k_2=3.5 \times 10^{-8} $ & Draine \& Sutin (1987), H12 \\
$\rm O(^3P_2) + H^+ \rightarrow O^+ + H$ & $k_3$~(Note a)& Stancil et al.\ (1999) \\
$\rm O^+ + H \rightarrow O + H^+$ & $k_4=5.7 \times 10^{-10} (T/{\rm 300\, K})^{0.36}\, e^{8.6 {\rm K}/T}$  & Stancil et al.\ (1999) \\
$\rm O^+ + H_2 \rightarrow OH^+ + H$ & $k_5=1.7 \times 10^{-9}$ & Smith et al.\ (1978) \\
$\rm OH^+ + H_2 \rightarrow H_2O^+ + H$ & $k_6=1.0 \times 10^{-9}$ &  Jones et al.\ (1981)\\
$\rm OH^+ + e \rightarrow O + H$ & $k_7=3.8 \times 10^{-8} (T/{\rm 300\, K})^{-0.5}$ & Mitchell (1990) \\

\enddata
\tablenotetext{a}{The rate coefficient $k_3$ is given by a more complex fitting function: see Stancil et al.\ (1999) for the expression adopted}

\end{deluxetable}

\noindent (1) For diffuse molecular clouds, the CRIR needed to match the observed H$_3^+$ abundances is an increasing function of $k_1$, the rate coefficient for dissociative recombination of H$_3^+$.  The dependence is linear in clouds where C is largely ionized.

\noindent (2) For diffuse atomic clouds, the CRIR needed to match the observed OH$^+$ abundances is an increasing function of $k_6$ and $k_7$, the rate coefficients for the dominant OH$^+$-destroying reactions.  

\noindent (3) For diffuse atomic clouds, the CRIR needed to match the observed OH$^+$ abundances and HRRL line strengths is linearly proportional to $k_2$, the rate coefficient for destruction of H$^+$ via charge transfer to PAHs, and inversely proportional to $k_5$, the rate coefficient for the production of OH$^+$ by the reaction of O$^+$ with H$_2$.

\noindent (4) For diffuse atomic clouds, the CRIR needed to match the observed OH$^+$ abundances is linearly proportional to $k_4/k_3$, i.e.\ the ratio of the rate coefficients for the destruction of O$^+$ via electron transfer from H to O$^+$ and for the formation of O$^+$ via electron transfer from O to H$^+$ and for 

In local thermodyamic equilibrium, $k_4/k_3$ is determined entirely by the principle of detailed balance and is a fixed function of temperature.  However, at the low densities of the interstellar clouds of present interest, atomic oxygen is almost entirely in the lowest fine structure state ($\rm ^3P_2$), with a negligible population in the excited states $\rm ^3P_1$ and $\rm ^3P_0$. This departure from LTE could significantly affect the value of $k_4/k_3$ if the channel to O($\rm ^3P_2$) is strongly disfavored in the reaction of O$^+$ with H.   The most widely adopted reaction rates for charge transfer involving  O and H$^+$, those computed by  Stancil et al.\ (1999; adopted in our study), do not show any such effect.  However, a subsequent theoretical study by Spirko et al.\ (2003) has suggested that the channel to O($\rm ^3P_2$) may indeed be abnormally slow at the temperatures of relevance; these authors cautioned, however, that the calculated cross-section for the reaction of O($\rm ^3P_2$) with H$^+$ is strongly dependent on the exact details of the assumed potential energy surface, and showed that minor modifications to the adopted potential could lead to large increases in that cross-section.  Both the Stancil et al.\ (1999) and Spirko et al.\ (2003) studies are consistent with laboratory measurements at 300~K that do not discriminate between O fine-structure states, so a definitive resolution of the issue must await future investigations.  While we have favored the Stancil et al.\ (1999) rate coefficients in our diffuse cloud models, we have investigated the effects of using those of Spirko et al.\ (2003) instead.  At 100~K and in the low-density limit (i.e.\ with all O in $\rm ^3P_2$), $k_4$ is decreased by a factor 1.3 relative to that of Stancil et al.\ (1999), while $k_3$ is decreased by a factor 5.8.  As a result, the value of $k_4/k_3$ is increased by a factor $\sim 4$, as is the CRIR required to match the observations.
If the Spirko et al.\ cross-sections are correct, then the CRIR estimated for diffuse atomic clouds becomes a factor $\sim 4$ larger than that for diffuse molecular clouds.

\subsection{Variation of the CRIR with Galactocentric radius}

\begin{figure}
\includegraphics[width=15 cm]{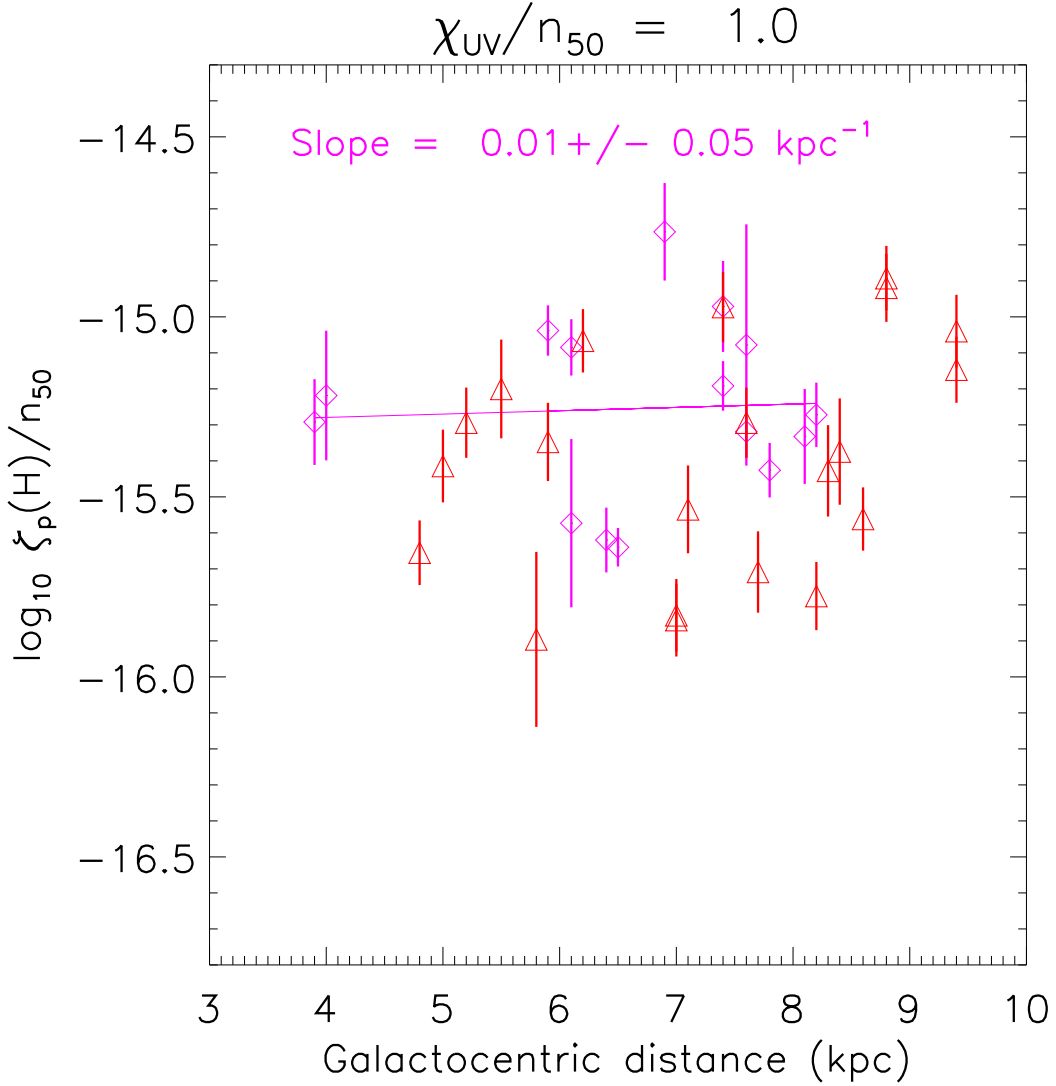}
\caption{Dependence of CRIR on Galactocentric radius, $R_g$, for diffuse atomic material in the Galactic disk.  Magenta diamonds: values of ${\rm log}_{10}[\zeta_p({\rm H})/n_{50}]$ determined from observations of OH$^+$, H$_2$O$^+$ and ArH$^+$.  Red squares: values of ${\rm log}_{10}[\zeta_p({\rm H})/n_{50}]$ determined from measurements of OH$^+$ and H$_2$O$^+$ alone. 
All values are computed for an assumed $\chi_{\rm UV}/n_{50}$ of 1.}
\end{figure}
 
Submillimeter observations of OH$^+$, $\rm H_2O^+$ and ArH$^+$ allow the CRIR to be determined for diffuse atomic material at considerably larger distances than is possible for the diffuse molecular clouds (observed with near-IR spectroscopy of H$_3^+$ and UV spectroscopy of H$_2$). As a result, we have obtained CRIR estimates for material covering a significant range of Galactocentric distances, $R_g$, from roughly 4 to 9~kpc. 
Following I15, we have used kinematic estimates of $R_g$ to examine the dependence of the CRIR in diffuse atomic clouds upon the Galactocentric distance.  

In Figure 13, we have plotted ${\rm log}_{10}[\zeta_p({\rm H})/n_{50}]$ versus $R_g$, with magenta diamonds showing CRIRs determined from measurements of OH$^+$, H$_2$O$^+$ and ArH$^+$, and red squares showing those determined from measurements of OH$^+$ and H$_2$O$^+$ alone.  All the estimates for $R_g$ are those given by I15.  For the red points, we adopted the mean correction factor needed to account for HI within the population of small clouds responsible for ArH$^+$ absorption.  Figure 13 indicates that there is no statistically-significant dependence of the derived values of  $\zeta_p({\rm H})/n_{50}$ upon $R_g$.  In particular, a linear fit to the (more reliable) magenta points yields a slope, $m$, of $0.01 \pm 0.05 \,{\rm kpc}^{-1}$; for the red points, the slope is $0.07 \pm 0.04 \,{\rm kpc}^{-1}.$    An important caveat here is that all the CRIRs plotted in Figure 13 were computed under the assumptions that $\chi_{\rm UV}/n_{50} = 1$ and $Z=Z_{\rm std}$.  Although the $\zeta_p({\rm H})/n_{50}$ values that we derive under those assumptions show no dependence upon $R_g$, additional considerations allow a Galactocentric dependence of $\zeta_p({\rm H})$ to be inferred as described below.

From the sensitivity analysis described in \S 4.3 above, we know that the derived values of $\zeta_p({\rm H})/n_{50}$ are roughly proportional to $[\chi_{\rm UV}/n_{50}]^{-0.7}$ and almost independent of $Z/Z_{\rm std}$.  We may therefore estimate the true Galactocentric gradient in ${\rm log}_{10}[\zeta_p({\rm H})/n_{50}]$ as
$${d {\rm log}_{10} [\zeta_p({\rm H})/n_{50}] \over dR_g} = m + 0.7\, {d {\rm log}_{10} n_{50} \over dR_g} - 0.7\,{d {\rm log}_{10} \chi_{\rm UV} \over dR_g},\eqno(7)$$
or equivalently,
$$ {d {\rm log}_{10} \zeta_p({\rm H}) \over dR_g} = m + 1.7\, {d {\rm log}_{10} n_{50} \over dR_g} - 0.7\,{d {\rm log}_{10} \chi_{\rm UV} \over dR_g}. \eqno(8)$$

Wolfire et al.\ (2003) have presented a comprehensive model of the neutral ISM within the Galactic disk, in which the UV radiation field has a scale length $-(d {\rm ln} \chi_{\rm UV}/dR_g)^{-1} = 4.1\, {\rm kpc}^{-1}$; this then implies a value of $-0.106$ for $d{\rm log}_{10} \chi_{\rm UV}/dR_g.$  In this model, the mean density in the cold neutral medium has an average Galactocentric gradient $d{\rm log}_{10} n_{\rm 50}/dR_g = -0.110 \, {\rm kpc}^{-1}$ (fit to Wolfire et al.\ 2003, Table 3, for $R_g$ in the range 3 to 8.5~kpc).  Thus, with the aid of equation (7), we obtain a best estimate the true Galactocentric gradient in the CRIR as
$$ {d {\rm log}_{10} \zeta_p({\rm H}) \over dR_g} = 0.01 - 1.7 \times 0.106 + 0.7 \times 0.111 = 0.093,  \eqno(9)$$
corresponding to a scale length $R_{\zeta} = -(d {\rm ln} \zeta_p({\rm H})/dR_g)^{-1} = 4.7\,{\rm kpc}.$

In the Wolfire et al.\ (2003) model, the mean density in the cold neutral medium is $n_{\rm H} = 33 \rm \, cm^{-3}$ at the solar circle ($R_g = R_0 = 8.5$~kpc).  Combining this density estimate with the mean CRIR listed in Table 2 and the Galactocentric radius dependence discussed above, we obtain the following estimate of the mean CRIR in diffuse atomic clouds
$$\zeta_p({\rm H}) = (2.2 \pm 0.3) \exp [(R_0-R_g)/4.7\,\rm{kpc}] \times 10^{-16} \rm \, s^{-1}.\eqno(10)$$
This value is entirely consistent with the mean CRIR determined from H$_3^+$ measurements in diffuse {\it molecular} clouds near the solar circle, $\zeta_p({\rm H}) = (2.3 \pm 0.6) \times 10^{-16} \rm \, s^{-1}$, and provides no 
evidence for any difference between the CRIR in diffuse atomic material and in diffuse molecular clouds.  By contrast, there is strong evidence (e.g.\ I15) that the CRIR is smaller by an order of magnitude or more in dense molecular clouds than it is in the diffuse ISM.

\section{Summary}

We have obtained estimates for the cosmic-ray ionization rate (CRIR) in the Galactic disk, using a detailed model for the physics and chemistry of diffuse interstellar gas clouds to interpret previously-published measurements of the abundance of four molecular ions: ArH$^+$, OH$^+$, $\rm H_2O^+$ and $\rm H_3^+$. 

\noindent (1) Within diffuse atomic clouds observed along the sightlines to bright submillimeter continuum sources, measurements of ArH$^+$, OH$^+$, $\rm H_2O^+$, and H column densities imply a mean logarithm of the 
CRIR of $ < \log_{10}[\zeta_{\rm p}({\rm H})/n_{\rm 50}] > \,\, = -15.34 \pm 0.05$, 
corresponding to a CRIR of $(4.6 \pm 0.5) \times 10^{-16}\, n_{\rm 50}\,\rm s^{-1}$, where $\zeta_{\rm p}({\rm H})\, {\rm s}^{-1} \sim [\zeta_{\rm t}({\rm H})/1.5] \, {\rm s}^{-1}$ is the primary ionization rate per H atom, $\zeta_{\rm t}({\rm H})\, {\rm s}^{-1}$ is the total ionization rate per H atom, $50\,n_{\rm 50}\, {\rm cm}^{-3}$ is the density of H nuclei, and the quoted errors are standard errors on the mean.  These CRIR estimates were obtained under the assumption that $\chi_{UV} /n_{50} = 1$, where $\chi_{UV}$ is the adopted UV radiation field in units of the mean value at the solar circle; they scale roughly as $(\chi_{UV} /n_{50})^{-0.7}$. 

\noindent (2) Within diffuse atomic clouds, the intrinsic dispersion of 
$\log_{10}[\zeta_{\rm p}({\rm H})/n_{\rm 50}]$ is estimated as 0.23, corresponding to a factor 1.7.

\noindent (3) Given existing models for the variation of mean gas density and UV radiation field with position within the Galactic disk, our analysis of the ArH$^+$, OH$^+$, and $\rm H_2O^+$ data leads to a recommended value of $\zeta_{\rm p}({\rm H})= (2.2 \pm 0.3) \exp [(R_0-R_g)/4.7\,\rm{kpc}] \times 10^{-16} \rm \, s^{-1}$ for the mean CRIR at Galactocentric distance $R_g$, where $R_0=8.5$~kpc.  

\noindent (4) Within diffuse molecular clouds observed toward stars in the solar neighborhood, measurements of $\rm H_3^+$ and $\rm H_2$ imply a mean logarithm of the CRIR of $ < \log_{10}\,\zeta_{\rm p}({\rm H}) > \,\, = -15.63 \pm 0.10$, 
corresponding to a CRIR of $(2.3 \pm 0.6) \times 10^{-16}\,\,\rm s^{-1}$ and a total ionization rate per H$_2$ molecule of $\zeta_{\rm t}({\rm H_2}) \sim 2.3 \,\zeta_{\rm p}({\rm H}) = (5.3 \pm 1.1) \times 10^{-16}\,\,\rm s^{-1},$ in good accord with previous estimates (IM12).

\noindent (5) For diffuse molecular clouds in which H$_3^+$ is detected, the intrinsic 
dispersion of $\log_{10}\,\zeta_{\rm p}({\rm H})$ is estimated as 0.09, corresponding to a factor of only 1.23. observations of H$_3^+$.   

\noindent (6) Our results show marginal evidence that the CRIR in diffuse molecular clouds decreases with cloud extinction, with a best-fit dependence $\propto A_{\rm V}({\rm tot})^{-1}$ for $A_{\rm V}({\rm tot}) \ge 0.5$.

\noindent (7) We have presented a rederivation of the CRIR implied by recent observations of carbon and hydrogen radio recombination lines along the sight-line to Cas A, which yields a best-fit estimate for the primary CRIR of $2.9 \times 10^{-16}\,\,\rm s^{-1}$ per H atom.  

\noindent (8) The uncertainty estimates presented in this paper are statistical in nature and do not include systematic uncertainties inherent in the diffuse cloud models.  We have identified several key reactions with rate coefficients that are important in determining what CRIR is needed to match the astronomical data: these include the dissociative recombination of H$_3^+$ and OH$^+$, the H abstraction reactions of O$^+$ and OH$^+$ with H$_2$, and the charge transfer reactions of H with O$^+$ and of O($\rm ^3P_2$) with H$^+$.  While our model adopts rate coefficients for these processes that are based upon the theoretical and experimental data currently available, we anticipate that new calculations and experiments may require them to be revised in the coming years; accordingly, we have discussed the dependences of the derived CRIRs upon the adopted rate coefficients for each of these processes.

\begin{acknowledgements}
We thank N.\ Indriolo and J.\ Black for several valuable comments about an earlier draft of this paper.  We gratefully acknowledge the support of grant number 120364 from NASA's Astrophysical Data Analysis Program (ADAP; NNX15AM94G). 

\end{acknowledgements}

\end{document}